\documentclass[twocolumn]{aastex63}

\usepackage{xspace}
\usepackage{amsmath}
\usepackage{threeparttable}
\usepackage{threeparttablex}
\usepackage{hyperref}
\usepackage{listings}
\usepackage{graphicx}
\usepackage{epsfig}
\usepackage{tabularx}
\usepackage{float}
\usepackage{diagbox}

\newcommand{\Tsys}{$T_{\mathrm{sys}}$\xspace}
\newcommand{\Trx}{$T_{\mathrm{rx}}$\xspace}
\newcommand{\Tsky}{$T_{\mathrm{sky}}$\xspace}

\newcommand{\Twvr}{$T_{\mathrm{WVR}}$\xspace}
\newcommand{\TsysNorm}{$\hat{T}_{\mathrm{sys}}$\xspace}
\newcommand{\TwvrNorm}{$\hat{T}_{\mathrm{WVR}}$\xspace}

\accepted{Nov. 22, 2022}
\shorttitle{ALMA Tsys tracking with WVR Data}
\shortauthors{He et al.}

\begin{document}

\title{Tracking ALMA System Temperature with Water Vapor Data at High Frequency}

\correspondingauthor{Hao He}
\email{heh15@mcmaster.ca}

\author[0000-0001-9020-1858]{Hao He}
\affiliation{McMaster University \\
	1280 Main St W, Hamilton, ON L8S 4L8, CAN}

\author{William R. F. Dent}
\affiliation{Joint ALMA Observatory \\
Alonso de C\'ordova 3107, Vitacura , Santiago, Chile}

\author{Christine Wilson}
\affiliation{McMaster University \\
	1280 Main St W, Hamilton, ON L8S 4L8, CAN}



\begin{abstract}

As the world-leading submillimeter telescope, the ALMA observatory is now putting more focus on high-frequency observations at Band 7 -- 10 (frequencies from 275 -- 950 GHz). However, high-frequency observations often suffer from rapid variations in atmospheric opacity that directly affect the system temperature $T_{\mathrm{sys}}$. Current observations perform discrete atmospheric calibrations (Atm-cals) every few minutes, with typically 10 -- 20 occurring per hour for high frequency observation and each taking 30 -- 40 seconds. In order to obtain more accurate flux measurements and reduce the number of atmospheric calibrations (Atm-cals), a new method to monitor $T_{\mathrm{sys}}$ continuously is proposed using existing data in the measurement set. In this work, we demonstrate the viability of using water vapor radiometer (WVR) data to track the $T_{\mathrm{sys}}$ continuously. We find a tight linear correlation between $T_{\mathrm{sys}}$ measured using the traditional method and $T_{\mathrm{sys}}$ extrapolated based on WVR data with scatter of 0.5\% -- 3\%. Although the exact form of the linear relation varies among different data sets and spectral windows, we can use a small number of discrete $T_{\mathrm{sys}}$ measurements to fit the linear relation and use this heuristic relationship to derive $T_{\mathrm{sys}}$ every 10 seconds. Furthermore, we successfully reproduce the observed correlation using atmospheric transmission at microwave (ATM) modeling and demonstrate the viability of a more general method to directly derive the $T_{\mathrm{sys}}$ from the modeling. We apply the semi-continuous $T_{\mathrm{sys}}$ from heuristic fitting on a few data sets from Band 7 to Band 10 and compare the flux measured using these methods. We find the discrete and continuous $T_{\mathrm{sys}}$ methods give us consistent flux measurements with differences up to 5\%. Furthermore, this method has significantly reduced the flux uncertainty due to \Tsys variability for one dataset, which has large precipitable water vapor (PWV) fluctuation, from 10\% to 0.7\%.  
\end{abstract}




\section{Introduction} \label{sec:intro}

\subsection{Flux Calibration in ALMA}

Calibration is the process by which the astronomer converts electronic signals from the telescope into meaningful astronomical data. Accurate calibration is crucial for the Atacama Large Millimeter/submillimeter Array (ALMA), as millimeter and submillimeter wavelength radiation will be adversely affected by the atmosphere and the electronic signal path in a variety of ways, and the antennas will also be affected by the observing environment \citep[ALMA technical handbook, Chapter 10]{Remjian_2019}. One of the important calibration processes is the amplitude and flux calibration. The aim of this calibration is to convert the raw visibilities (and auto-correlations) from the correlator into brightness temperature or flux density by carefully tracking the instrumental and atmospheric variations and determining accurate conversion factors. Because of the large and rapidly varying opacity of water vapor, standard calibration procedures are less accurate at submillimeter wavelengths. For the flux calibration, a well defined scientific goal can be elucidated and set as the requirement. In numerous meetings and discussions, the scientific community originally made clear its desire to reach 1\% flux density accuracy \citep[e.g.][report of the spring 2003 ASAC meeting]{ASAC_2003}, which means that we must be able to determine the overall flux density scale (and apply it to the visibilities and total power measurements) to 1\% accuracy. In addition, the capability of achieving a dynamic range of 10000 or higher in ALMA images means that we must track the amplitude fluctuations to better than 1\% \citep{Yun_211}. A later study by \citet{Moreno_memo372} showed that it is impractical to achieve 1\% at submillimeter wavelengths, and so a requirement of 3\% has been adopted for frequencies $>$300 GHz. The current achieved calibration accuracy for ALMA is 5\% at the lower bands (100 GHz), 10\% in mid-bands (200--400 GHz) and 20\% in the higher bands ($>$400 GHz) \citep{Remjian_2019}. This paper is part of the work being done to improve the overall flux calibration accuracy.  

Currently there are two flux calibration strategies, the astronomical flux calibration and the direct instrumental amplitude calibration. The astronomical flux calibration method uses an astronomical source with known flux and scales up the recorded amplitude based on that flux standard. This method requires the astronomical source to be bright and have stable flux. Currently planets are used as the primary flux standard while quasars are also used as an alternative flux standard due to their compact size and availability over the sky. However, the estimated accuracy of the planets' fluxes is only about 10\% \citep[10.4.7]{Remjian_2019}. Therefore, people are still searching for ideal flux calibrators with high accuracy, especially at high frequencies. In addition to the flux calibrator standard, the relative sky transmission on the calibrator and the science target, as well as the time variability of the transmissions will also affect the overall flux calibration. This is the standard method currently used by ALMA. On the other hand, for a stable system, one can directly translate the measured counts in total power into flux units using a direct instrumental amplitude calibration method. Both methods rely on accurate measurement of the sky opacity and tracking its variations during the observation. At millimeter wavelengths, the changes in atmospheric transparency will usually be very modest, under 1\% over 10 minutes about 80\% of the time. Since the same amount of water vapor results in much larger opacities in the submillimeter, the transparency fluctuations in the submillimeter over characteristic calibration time scales will be much larger, typically several percent during median stability conditions and sometimes $>$ 10\%. 

For ALMA, both calibration methods require the precise measurement of the system temperature \Tsys and complex gain $G$. \Tsys represents the total thermal noise of the measurement. \Tsys includes contributions from the sky, receiver, and system losses, with a large contribution coming from the sky temperature. Since ALMA is equipped with receivers of sufficiently low noise, the sky noise often dominates the total thermal noise. Therefore, it is necessary to track the changes in system temperatures caused by the fluctuations in the atmosphere. Current ALMA \Tsys measurements use discrete atmosphere (ATM) calibrations done every few minutes with a cadence depending on the observing band. At low frequencies ($<$ 300 GHz), ALMA generally perform 2 or 3 \Tsys measurements over a typical hour-long observation due to the assumed small variation in the atmosphere transmission. At high frequencies ($>$ 300 GHz), due to the rapid opacity change in atmosphere, ALMA generally perform 10 $\sim$ 20 ATM calibrations per hour. So the time overheads just due to ATM calibration can become quite significant -- up to 15-20\% at the highest bands (9 and 10, at 602--950 GHz). Moreover the variations of \Tsys on timescales faster than the ATM calibration interval are not tracked with this discrete ATM calibration method.  Therefore, one of the major goals in high-frequency flux calibration is to track \Tsys more closely while also reducing the time spent on discrete ATM calibrations.

\subsection{\Tsys Measurements in Flux Calibration}
The system temperature (\Tsys) is the fundamental parameter to determine the system sensitivity and the real flux of the source.
Tsys includes various contributions, and can be written
in a basic form as \citep[adapted from ][]{Mangum_memo602}
\begin{equation} \label{eq:Tsys_def}
T_{\mathrm{sys}} = \frac{1}{\eta_{\mathrm{f}} e^{-\tau_{\mathrm{sky}}}} \left(T_{\mathrm{rx}} + \eta_{\mathrm{f}} T_{\mathrm{sky}} + (1-\eta_{\mathrm{f}}) \times T_{\mathrm{amb}} \right) 
\end{equation}
where 
\begin{itemize}
	\item $T_{\mathrm{rx}}$ is receiver temperature
	\item $T_{\mathrm{sky}}$ is sky temperature
	\item $T_{\mathrm{amb}}$ is ambient temperature where spillover is assumed to be terminated
	\item $\eta_{\mathrm{f}}$ is the forward efficiency. This is equal to the fraction of the antenna power pattern that is contained within the forward hemisphere and is currently assumed to be 0.95
	\item $e^{-\tau_{\mathrm{sky}}}$ is the fractional transmission of the atmosphere, where $\tau_{\mathrm{sky}}$ is equal to the atmospheric opacity along the target's line of sight. 
\end{itemize}
Note this equation is for single sideband (SSB) and sideband separating (2SB) receivers, which are used for ALMA Band 3 -- 8 observation. For this configuration, the image sideband gain is assumed negligibly small. The Band 9 and 10 receivers are using the double sideband configuration and hence \Tsys are calculated differently \citep[][ Eq. 6]{Mangum_memo602}. $T_{\mathrm{sky}}$ and $T_{\mathrm{amb}}$ can be further expressed as 
\begin{equation}
\label{eq:Tsky_Tamb}
\begin{split}
&T_{\mathrm{sky}} = T_{\mathrm{atm}} (1-e^{-\tau_{\mathrm{sky}}}) \\
&T_{\mathrm{amb}} \approx T_{\mathrm{atm}}
\end{split}
\end{equation}
where $T_{\mathrm{atm}}$ is the representative atmosphere temperature. Note that $T_{\mathrm{amb}} \approx T_{\mathrm{atm}}$ is a reasonable approximation when the opacity originates close to the ground (e.g. due to water vapor). Therefore, by combining equation \ref{eq:Tsys_def} and \ref{eq:Tsky_Tamb}, \Tsys can be calculated as
\begin{equation}
\begin{split}
T_{\mathrm{sys}} &= \frac{1}{\eta_{\mathrm{f}} e^{-\tau_\mathrm{sky}}} \left[T_{\mathrm{rx}}+T_{\mathrm{amb}} (1-\eta_{\text {f}} e^{-\tau_{\mathrm{sky}}})\right] \\ 
&\approx  \frac{1}{e^{-\tau_{\mathrm{sky}}}} \left(T_{\mathrm{rx}}+T_{\mathrm{sky}} \right) \: (\mathrm{where}\  \eta_{\text{f}} \sim 1)
\end{split}
\label{eq:Tsys_simplified}
\end{equation} 
This equation suggests that the key parameters to measure \Tsys are \Trx and \Tsky (as $\tau_{\mathrm{sky}}$ can be derived from \Tsky using Eq. \ref{eq:Tsky_Tamb}). 

In ALMA flux calibration, the intensity of the observed source is directly proportional to \Tsys by the following equation \citep[e.g.][]{Brogan_2018}
\begin{equation}
\label{eq: signal_Tsys}
S_{\mathrm{final}} \sim S_0 \times \sqrt{T_{\mathrm{sys}}(i)T_{\mathrm{sys}}(j)} \times \Gamma 
\end{equation}
where $S_0$ and $S_{\mathrm{final}}$ are the fluxes measured before and after the flux calibration. $\Gamma$ is the antenna efficiency factor to convert K to Jy and $i$ and $j$ represent the two antennas forming the baseline. Note that ALMA uses an antenna-based calibration method to simplify the calibration process. Nearly all of the changes to the visibility function (e.g. atmosphere, system noise, amplitude changes, delay changes) can be decomposed into the two complex antenna-based gain factors associated with any baseline. This approach reduces the number of gain correction terms for an N-element array from $N(N-1)/2$ baselines to $N$ antennas. In this case, \Tsys is associated with each antenna and $S_{\mathrm{final}}$ is associated with each baseline. 

\Tsys also determines the achieved rms noise of the observation \citep[e.g.][a modified form of radiometer equation]{Condon_Ransom_2016}
\begin{equation}
\mathrm{rms} \approx \frac{c T_{\mathrm{sys}}}{\sqrt{\Delta \nu t_{\mathrm{int}}}}
\end{equation}
where $\Delta \nu$ is the frequency bandwidth, $t_{\mathrm{int}}$ is the integration time of the observation and $c$ includes the quantization and correlator efficiencies, and is typically 0.8-0.9 for ALMA. Therefore, higher \Tsys means the data has a larger noise within one observation. In addition, for ALMA the weighting function used to combine visibility data is inversely proportional to \Tsys as
\begin{equation}
\mathrm{Weight} \propto \frac{1}{T_{\mathrm{sys}}(i)T_{\mathrm{sys}}(j)}
\end{equation}

\subsection{Traditional Method to Measure \Tsys}

As noted above, \Tsys is highly dependent on the sky opacity (eq. \ref{eq:Tsky_Tamb}). ALMA antennas use a two-load system for \Tsys measurement in Band 3 and higher, which is different from the one-load system used for other radio telescopes. The two-load system in theory can achieve a \Tsys measurement accuracy of 1\%, which is significantly better than that of a one-load system ("chopper wheel") of 5\% \citep{Yun_211}. For ALMA, \Tsys is obtained from an atmospheric calibration (ATM-cal) scan where a hot load, ambient load and sky are consecutively placed in front of the feed using an Amplitude Calibration Device \citep[ACD;][]{Casalta_2008}. Typically this process takes 30-40 seconds, including antenna slew time and overheads. At frequencies below about 400 GHz, where the system temperatures are more stable (except in the 183 and 325 GHz water lines), an ATM-cal scan is made every 10 to 20 minutes. However, at higher frequencies, and wherever the opacity is large and more variable, every scan on the astronomical target will have an associated ATM-cal measurement \citep{Remjian_2019}, implying a cadence of ATM calibration as fast as once every 2-3 minutes. From the two-load system, we can also measure \Trx. In this case, the measured power can be expressed as \citep{Mangum_434}
\begin{equation}
\begin{split}
P_{\mathrm{hot}} & = K \  (T_{\mathrm{rx}}+T_{\mathrm{hot}})  \\
P_{\mathrm{amb}} &=  K \  (T_{\mathrm{rx}}+T_{\mathrm{amb}}) \\
P_{\mathrm{sky}} &= K \ (T_{\mathrm{rx}}+T_{\mathrm{sky}})
\end{split}
\end{equation}
where $K$ is the gain to convert the temperature to the measured power. $T_{\mathrm{hot}}$ and $T_{\mathrm{amb}}$ are generally about 350 K and 290 K. Based on the equation above, we can express \Trx and \Tsky as 
\begin{equation}
\begin{split}
T_{\mathrm{rx}} &= \frac{T_{\mathrm{hot}}P_{\mathrm{amb}}-T_{\mathrm{amb}}P_{\mathrm{hot}}}{P_{\mathrm{hot}}-P_{\mathrm{amb}}} \\
&= \frac{T_{\mathrm{hot}} - Y_1 T_{\mathrm{amb}}}{Y_1 -1} \\
T_{\mathrm{sky}} &= \frac{P_{\mathrm{sky}}T_{\mathrm{amb}}-(P_{\mathrm{amb}}-P_{\mathrm{sky}})T_{\mathrm{rx}}}{P_{\mathrm{amb}}} \\
&= Y_2 T_{\mathrm{amb}} - (1-Y_2) T_{\mathrm{rx}}
\end{split}
\end{equation}
where $Y_1 \equiv P_{\mathrm{hot}} / P_{\mathrm{amb}}$ and $Y_2 \equiv P_{\mathrm{sky}} / P_{\mathrm{amb}}$. Unlike the atmosphere, \Trx is relatively constant throughout the observation. Measurements performed by ALMA show fluctuations of \Trx are generally smaller than 1\% during normal sidereal tracking. With the measurement of \Tsky, we can further derive the optical depth based on Eq. \ref{eq:Tsky_Tamb} and calculate the \Tsys based on Eq. \ref{eq:Tsys_simplified}. In summary, the expressions for the key quantities to measure \Tsys are 
\begin{equation}
\label{eq:Tsys_Tsky_Trx}
\begin{split}
T_{\mathrm{rx}} &= \frac{T_{\mathrm{hot}} - Y_1 T_{\mathrm{amb}}}{Y_1 -1} \approx \text{const} \\
T_{\mathrm{sky}} &=  Y_2 T_{\mathrm{amb}} - (1-Y_2) T_{\mathrm{rx}} \\
T_{\mathrm{sys}} & \approx \frac{1}{e^{-\tau_{\mathrm{sky}}}} \left(T_{\mathrm{rx}}+T_{\mathrm{sky}} \right)
\end{split}
\end{equation}
Therefore, during each ATM cal, we point the array to hot load, ambient load and sky to measure \Trx and \Tsky and then calculate the \Tsys at that time.  


\subsection{Candidate Data to Track the Continuous \Tsys}
\label{sec:candidate_data}

As mentioned above, the current method takes extra time to obtain a spot measurement of \Tsys every few minutes. If we want to continuously track \Tsys, in theory there are 3 types of measurement data available from ALMA to achieve this goal: Water Vapor Radiometer (WVR) data, auto-correlation (AC) data or square law detector (SQLD) data. We will describe where these data arise, and the theory behind each method below. The advantages and disadvantages of each method are summarized in Table. \ref{tab:Tsys_tracking}.

\begin{table*}[]
\renewcommand{\arraystretch}{1.2}
\begin{threeparttable}
\centering
\movetableright=-0.5 in
\caption{Summary of different methods to track \Tsys}
\label{tab:Tsys_tracking}
\begin{tabular}{@{}cll@{}}
	\hline
	\hline
	& \multicolumn{1}{c}{WVR$^1$}                                          & \multicolumn{1}{c}{AC \& SQLD$^2$}             \\
	\hline
	Advantage            & 1.  The data is continuously calibrated  to measure ${T_{\mathrm{WVR}}}^{\mathrm{a}}$      & 1. Directly proportional to \Tsys when $\tau_{\mathrm{sky}}$ is small               \\
	\multicolumn{1}{l}{} & 2. Does not suffer from internal electronic gain drift$^{\mathrm{b}}$                 & 2. Has the same frequency coverage as the science target \\
	Disadvantage         & 1. Has different frequency coverage as the science target        & 1. The data is not calibrated.                   \\
	& 2. Directly tracks \Tsky not \Tsys & 2. Suffers from the electronic gain drift$^{\mathrm{b}}$ or gain variations \\
	& & 3. For AC, no linearity correction in FDM mode.  \\
	\hline
\end{tabular}
\begin{tablenotes}
\item{\textbf{Columns:}} 1. Water vapor radiometer data.  2. Auto-correlation and square law detector data. 
\item \textbf{References:} a. \citet{Hills_memo352}. b. \citet{Payne_2001}.
\end{tablenotes}
\end{threeparttable}
\end{table*}

\begin{figure*}[htb!]
	\gridline{
		\fig{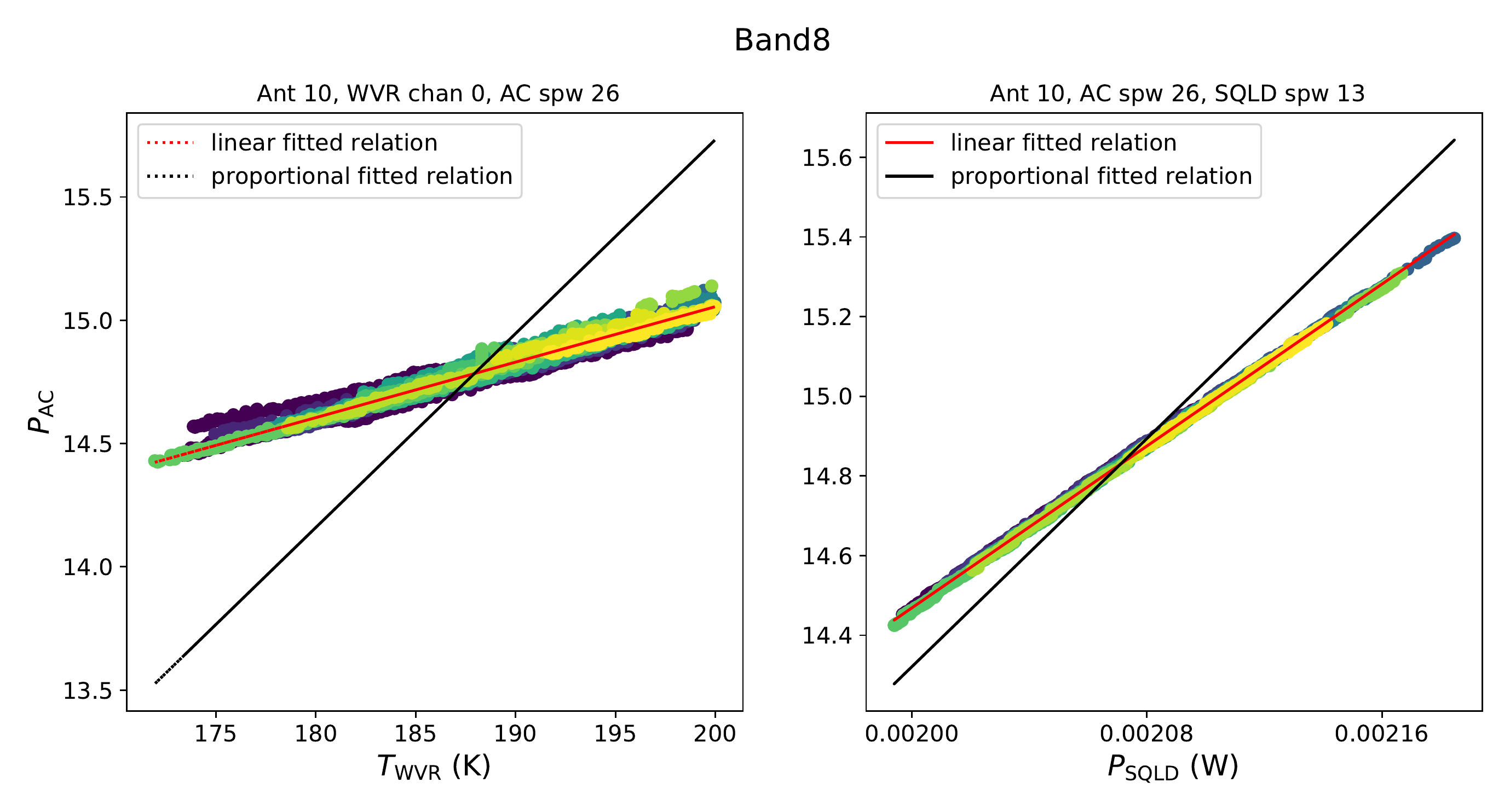}{0.8\linewidth}{}
	} 
\vspace{-2.5\baselineskip}
	\caption{The correlation between water vapor radiometer (WVR) data, auto-correlation (AC) data and square law detector (SQLD) data for antenna 10 for dataset Band8 data. The scatter plots are color coded by the scan number. (Left) AC data versus WVR data. The red line is the best linear fit and black line is the proportional fit through 0. We have excluded WVR data greater than 200 as those are from samples taken on the hot load and ambient load. (Right) AC data versus SQLD data. }
	\label{fig: WVR_TP_SQLD}
\end{figure*} 

The WVR data are used by ALMA to track the optical depth of the water vapor along the line of sight to each antenna, and hence are used to correct for the resulting effective pathlength and delay errors. The WVRs do Dicke switching and have internal calibrated loads, so the output from each WVR is the calibrated sky temperature (\Twvr) at 4 frequencies \citep[184.19, 185.25, 186.485 and 188.51 GHz respectively;][]{Hills_memo495} around the 183 GHz water line taken every 1.152 seconds \citep[Section A.6]{Remjian_2019}. By comparing \Twvr with $T_{\mathrm{amb}}$, we can calculate the precipitable water vapour (PWV), which is proportional to the atmospheric opacity caused by the water absorption. Since \Tsky at our observing frequency (Eq. \ref{eq:Tsys_Tsky_Trx}) and \Twvr are tracking sky temperatures at different frequencies, we would expect
\begin{equation}
\label{eq:Tsky_Twvr_tau}
\begin{split}
\tau_{\mathrm{sky}} &= C \times \tau_{\mathrm{WVR}} + \tau_{\mathrm{dry}} \\
T_{\mathrm{sky}} &= C \times T_{\mathrm{WVR}} + T_{\mathrm{sky, dry}} \quad (\tau_{\mathrm{sky}}, \tau_{\mathrm{WVR}} << 1)
\end{split}
\end{equation}
where $\tau_{\mathrm{sky}}$ is the sky opacity at the observed frequencies , $\tau_{\mathrm{WVR}}$ is the optical depth at the WVR channel frequency, $\tau_{\mathrm{dry}}$ and $T_{\mathrm{sky, dry}}$ are the optical depth and sky temperature contribution for the dry component at the observing frequency, and $C$ is a constant. A small dry contribution to $\tau_{\mathrm{WVR}}$ and $T_{\mathrm{WVR}}$ is not explicitly shown but does not change the form of the relationship. The overall sky opacity includes contributions from the wet component (H$_2$O lines from the troposphere which are relatively wide due to pressure broadening), and from the dry component (mostly due to lines of O$_2$ and O$_3$, but also including a continuum component as well as other molecules).  If optical depth is small enough, we would expect the observed temperature is proportional to the optical depth and hence the proportion relation between the two optical depths holds also for the two measured temperatures. 
Since the major change in \Tsys is caused by the variation in \Tsky, we would expect that \Twvr is tracking \Tsys. The major advantage of using the WVR data to trace \Tsys is that the radiometer is constantly monitoring the sky and internally calibrating itself. Therefore, we can extrapolate \Tsys throughout the entire observation based on the WVR data. Furthermore, since \Twvr is internally calibrating and tracking the sky variation, it does not suffer from the internal electronic drift or small changes in system gain, which can affect the measured values of an uncalibrated signal (see Table \ref{tab:Tsys_tracking}). 

Alternatively, we would expect \Tsys is tightly correlated with the total power signal received by each antenna. To be more precise, the total power signal should be directly proportional to ($T_{\mathrm{rx}}+T_{\mathrm{sky}}$), which can be used to calculate \Tsys given the optical depth $\tau_{\mathrm{sky}}$ using Eq. \ref{eq:Tsys_Tsky_Trx} (see detailed discussion in Section \ref{sec:AC}).  
The total power signal received by each antenna is measured by a square-law detector (SQLD) built into the ALMA signal path, whose data is also recorded in the datasets. Additionally, the autocorrelation data recorded in the measurement set should also give us the total signal received by each antenna. We would expect
\begin{equation}
T_{\mathrm{rx}}+T_{\mathrm{sky}} \propto P\mathrm{_{AC}} \propto P\mathrm{_{SQLD}}
\end{equation}
where $P\mathrm{_{AC}}$ and $P\mathrm{_{SQLD}}$ are the power of auto-correlation data and SQLD data read from the measurement set, respectively. If $\tau_{\mathrm{sky}}$ is small, we would expect direct proportionality between \Tsys and the total power received which could help us derive continuous \Tsys. In addition, both AC and SQLD data cover the same frequency rangels as the actual observed science data so we do not need to assume atmosphere variation has the same effect on data at different wavelengths (the constant $C$ in Eq. \ref{eq:Tsky_Twvr_tau} and the explicit dry contributions). 

In Fig. \ref{fig: WVR_TP_SQLD}, we plot the correlation between AC and WVR and SQLD data. We can see the AC and SQLD data follow a tighter linear correlation. These two types of data are expected to be equivalent and thus should follow a proportional correlation. We see an offset from direct proportionality between AC and SQLD data in this observation as no linearity correction for the effect of the 3 bit samplers is applied in this correlator mode, and there can be residual DC offsets in the SQLD data. On the other hand, we can see that the WVR and AC data do not follow the same proportional relation. This can be caused by various reasons summarized in Table \ref{tab:Tsys_tracking}. In particular, the distribution in the AC data at similar WVR levels on the left panel is indicative of slightly different system gains in different scans during the observation, or the differing wet and dry opacity contributions at different aimasses. 
In this case, we need to compare the two types of data to explore which one is better in tracking \Tsys. 

\subsection{Outline of This Paper}
\label{sec:outline}

In the following sections, we will explore how well different data track \Tsys measurements. In Section 2, we explore the viability of using \Twvr to track \Tsys. In Section 3, we use the Atmospheric Transmission at Microwave (ATM) modeling to test the theory behind the tight \Tsys vs \Twvr correlation. In Section 4, we explore the viability to use AC or SQLD data to track \Tsys. What we find is that those two types of data do not work well in tracking \Tsys. In Section 5, we describe our new calibration method to use alternative \Tsys derived from \Twvr and how it compares to the original discrete calibration method. 

For our analysis, we use measurement sets from several projects in Bands 7, 8, 9 and 10 \citep{Mahieu_2012, Sekimoto_2008, Baryshev_2015, Gonzalez_2014}. We also include two projects with multiple measurement sets from Band 7 and 9. The summary of the data we use is given in Table \ref{tab:data_summary}.  
\begin{table*}
\caption{Summary of Data}
\label{tab:data_summary}
\begin{threeparttable}
\begin{tabularx}{0.95\textwidth}{cclllcc}
	\hline \hline
	Dataset Label & Project        & \multicolumn{1}{c}{Band} & \multicolumn{1}{c}{Target}      & \multicolumn{1}{c}{Data uid} & PWV (mm) & Elev. (deg) \\ 
	\multicolumn{1}{c}{(1)} & (2) & \multicolumn{1}{c}{(3)} & \multicolumn{1}{c}{(4)} & \multicolumn{1}{c}{(5)} & \multicolumn{1}{c}{(6)} & (7) \\ \hline	
	Band10     & 2015.1.00271.S & \multicolumn{1}{c}{10}   & \multicolumn{1}{c}{Arp 220}     & uid://A002/Xbe0d4d/X12f5     & 0.28  &  43 \\
	Band9a     & 2016.1.00744.S & \multicolumn{1}{c}{9}    & \multicolumn{1}{c}{IRAS16293-B} & uid://A002/Xbf792a/X14cc     & 0.37  &  63 -- 76 \\
	Band8      & 2018.1.01778.S & \multicolumn{1}{c}{8}    & \multicolumn{1}{c}{SPT0311-58}  & uid://A002/Xdb7ab7/X1880b    & 0.85   & 53 -- 55 \\
	Band7a     & E2E8.1.00003.S & \multicolumn{1}{c}{7}    & \multicolumn{1}{c}{HT-Lup}      & uid://A002/Xec4ed2/X912      & 0.59   & 52 -- 66 \\
	Band7b1    & 2018.1.01210.S & \multicolumn{1}{c}{7}    & \multicolumn{1}{c}{AS205A}      & uid://A002/Xda1250/X2387     & 0.69   & 50 -- 64 \\
	Band7b2    &                &                          &                                 & uid://A002/Xda1250/X32df     & 0.53   &  80 -- 85 \\
	Band7b3    &                &                          &                                 & uid://A002/Xda845c/X35d1     & 0.5     & 55 -- 70 \\
	Band7b4    &                &                          &                                 & uid://A002/Xda1250/X3e39     & 0.51    & 60 -- 75 \\
	Band7b5    &                &                          &                                 & uid://A002/Xda1250/X4db3     & 0.49    & 35 -- 50 \\
	Band7b6    &                &                          &                                 & uid://A002/Xd99ff3/X15d7b    & 0.42    & 65 -- 80 \\
	Band7b7    &                &                          &                                 & uid://A002/Xd99ff3/X1702c    & 0.51    & 63 -- 77 \\
	Band7b8    &                &                          &                                 & uid://A002/Xd99ff3/X17da2    & 0.53    & 40 -- 55 \\
	Band9b1    & 2019.1.00013.S & \multicolumn{1}{c}{9}    & \multicolumn{1}{c}{Circinus}    & uid://A002/Xed9025/X769c     & 0.43    &  34 -- 42 \\
	Band9b2    &                &                          &                                 & uid://A002/Xed8123/X7b1      & 0.37    & 37 -- 43 \\
	Band9b3    &                &                          &                                 & uid://A002/Xed4607/X1208a    & 0.34    & 31 -- 39  \\ \hline
\end{tabularx}
\begin{tablenotes}[flushleft]
	\item \textbf{Columns:} (1) Label for each dataset used in this paper (2) ALMA project code (3) Observed Band. (4) The name of the science target to be observed (5) The ALMA Unique Identifier (UID) of each execution (6) The precipitable water vapor (PWV) column. (7) Elevation range of the science target 	
\end{tablenotes}
\end{threeparttable}
\end{table*}

\section{WVR Data to Track \Tsys}
\label{sec:WVR_Tsys}

In this section, we examine how well \Twvr tracks \Tsys and explore how the correlation is affected by various parameter choices. We mainly use the dataset Band8 (uid://A002/Xdb7ab7/X1880b) for illustration purposes. Examples from additional datasets are given in  Appendix \ref{sec:extra_figure}. 

\subsection{\Tsys versus \Twvr}
\label{sec:Tsys_Twvr}

To check whether \Tsys is tracked by the WVR data, we first need to match the WVR data taken at the same time as the \Tsys measurements. We then average the WVR values that are within 10 s around the time when \Tsys is measured and compare the averaged \Twvr with its corresponding \Tsys. 10 s is a typical time for one Atm-cal scan and hence is the shortest timescale we expect \Tsys to change. We also note that \Tsys recorded in the measurement set is a spectrum with two polarizations. In our analyses to compare \Tsys with \Twvr, we average \Tsys from both polarizations and also along the spectral axis within one spectral window (spw).

\begin{figure*}[htb!]
	\gridline{
		\fig{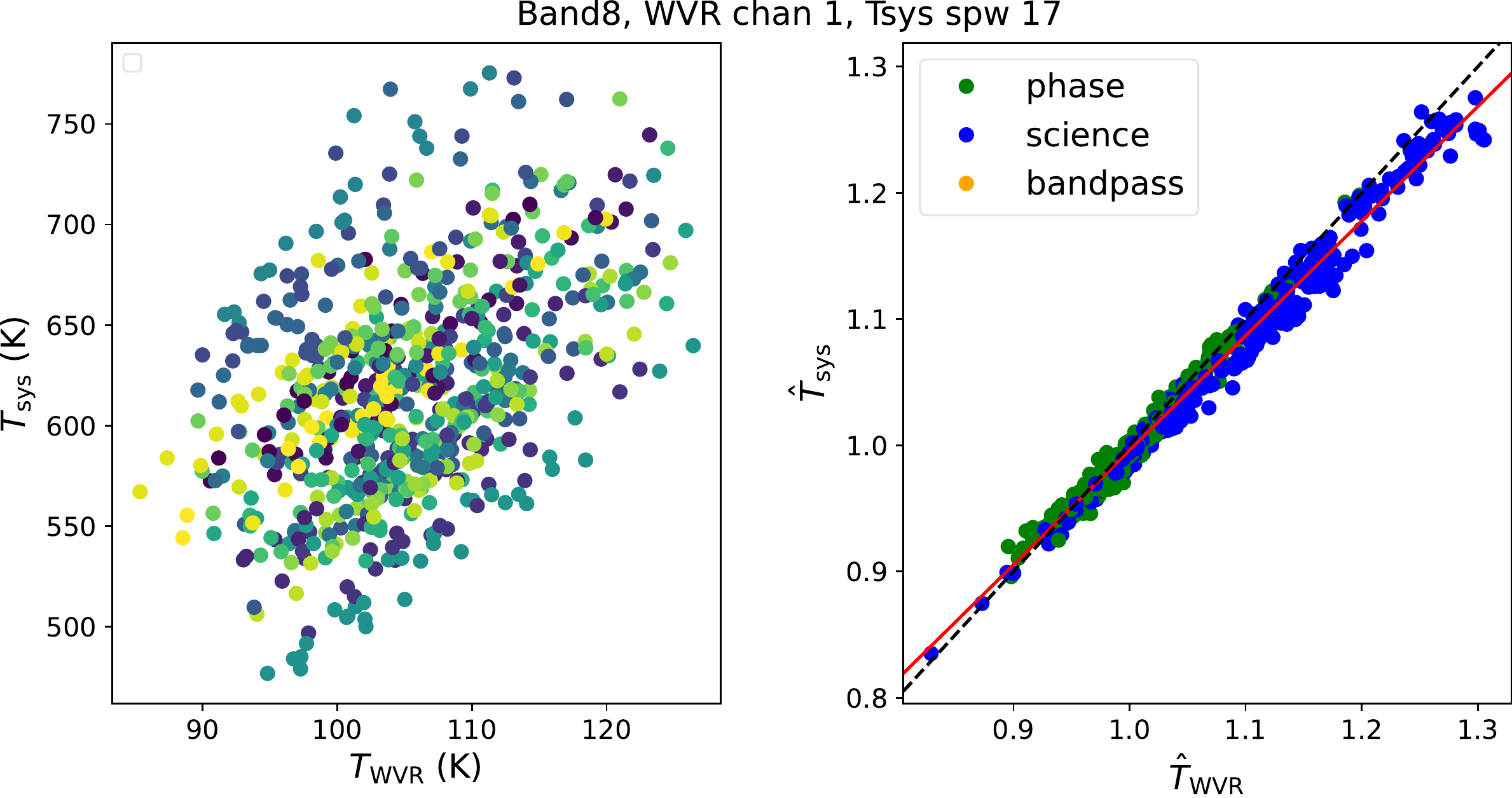}{0.8\linewidth}{}
	} 
	\vspace{-2\baselineskip}
	\caption{(Left) \Tsys vs \Twvr for dataset Band8 color coded by different antennas. (Right) \Tsys and \Twvr normalized to the value of first scan of each target (bandpass, phase, science) for each antenna. We can see the normalized \TsysNorm and \TwvrNorm follows a tight linear correlation. }
	\label{fig: Tsys_WVR_norm}
\end{figure*} 

We first plot \Tsys versus \Twvr from all antennas for each data set.  One example of \Tsys versus \Twvr is shown in the left panel of Fig. \ref{fig: Tsys_WVR_norm}. For this case, we select \Twvr from WVR channel 1. We will discuss in Section \ref{sec:wvr_chans} how the selection of different WVR channels affects the relation between \Twvr and \Tsys. As we can see, there is a significant correlation between \Tsys and \Twvr for each spw. However, the scatter is large along the direction perpendicular to the trend, as expected due to the differences in the receiver (\Trx and sideband gains) and WVR between antennas. Furthermore, since bandpass, phase-cal and science observations are observing targets at different elevations, it is possible that the scatter is also caused by data from different types of observations. Therefore, to see if the WVR tracks the time variation of \Tsys, we normalize \Tsys and \Twvr by the first  measurement for each observing target (bandpass, phase-cal and science) of each antenna as
\begin{equation}
\label{eq:Tsys_Twvr_norm}
\begin{split}
\hat{T}_{\mathrm{sys, source}} (t) &= \frac{T_{\mathrm{sys, obs}}(t)}{T_{\mathrm{sys, obs}}(1^{\mathrm{st}})} \\
\hat{T}_{\mathrm{WVR, obs}} (t) &= \frac{T_{\mathrm{WVR, obs}}(t)}{T_{\mathrm{WVR, obs}}(1^{\mathrm{st}})}
\end{split}
\end{equation}
where $\hat{T}_{\mathrm{sys}}$ and $\hat{T}_{\mathrm{WVR}}$ are the normalized values of \Tsys and \Twvr,  the subscript 'obs' is the generalized term for each type of observing target (bandpass, phase and science) and $1^{\mathrm{st}}$ in the bracket means the value when the first \Tsys for each observing target is measured. 

The correlation between the \TsysNorm and \TwvrNorm is shown in the right panel of Fig. \ref{fig: Tsys_WVR_norm}. We can see that these two variables have a tight linear correlation, with scatter less than 1\%. This tight linear correlation is also seen in other data sets, as illustrated in Appendix \ref{sec:extra_figure}. This indicates that \Twvr can be used to track the \Tsys if the slope and intercept can be determined for each spw or frequency. As described by eq. \ref{eq:Tsky_Twvr_tau}, 
the relation is expected to be frequency dependent, and it further differs from 1-to-1 due to the other contributions to \Tsys apart from \Tsky (e.g. Eq. \ref{eq:Tsys_def} and \ref{eq:Tsys_simplified}). In Section \ref{sec:ATM_modeling}, we explore the relationship using an atmospheric opacity model, but here we take a heuristic approach to determine the linear relationship from the data itself.

\begin{figure*}[tbh!]
	\gridline{
		\fig{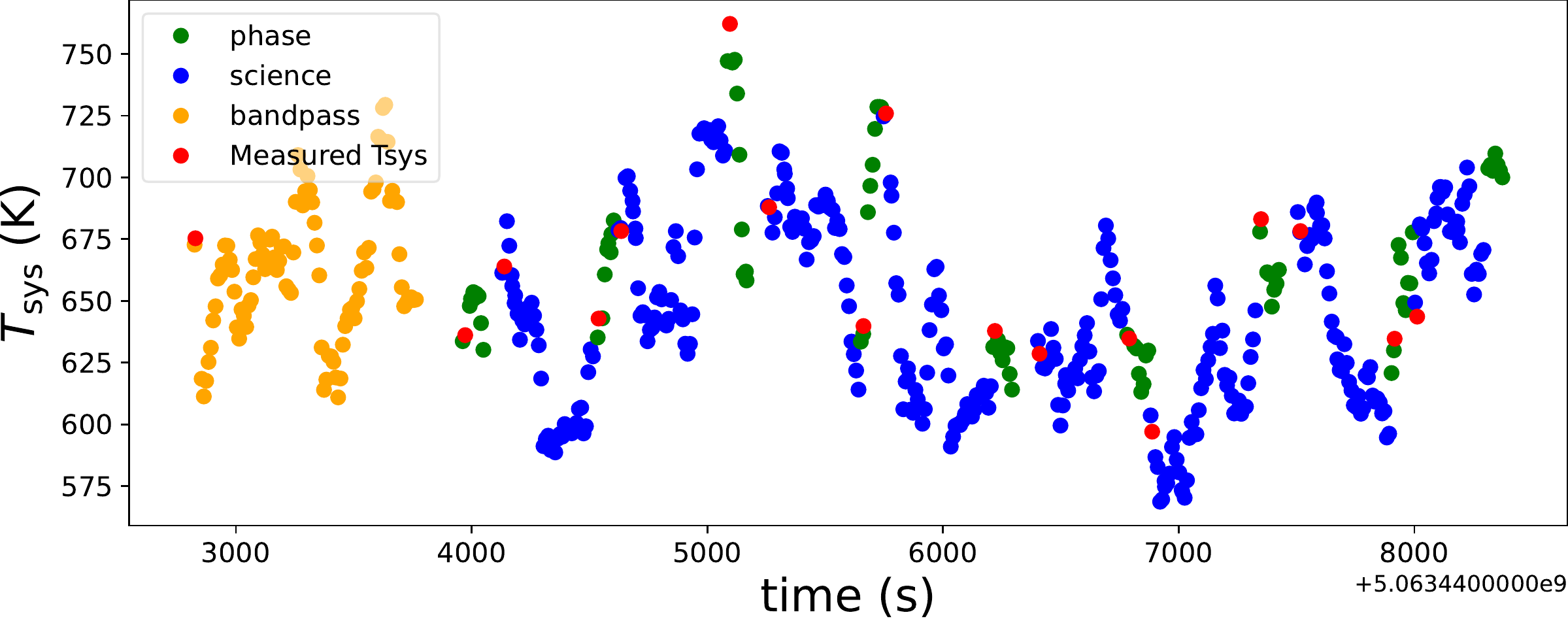}{0.8\linewidth}{}
	} 
	\vspace{-2\baselineskip}
	\caption{The extrapolated \Tsys versus the original \Tsys for antenna 10 of the dataset Band8 at spw 17. The orange, green and blue points are extrapolated continuous \Tsys for each observing target based on Eq. \ref{eq:Tsys_extrapolate} while the red points are original \Tsys measurements. }
	\label{fig:Tsys_WVR_extrap}
\end{figure*}

We can then use the fitted linear relation to extrapolate the continuous \Tsys based on the first \Tsys value for each observing target and the stream of \Twvr values. The exact equation can be expressed as 
\begin{equation}
\label{eq:Tsys_extrapolate}
\begin{split}
T_{\mathrm{sys}}(t) &= T_{\mathrm{sys}}(\mathrm{1^{st}}) \cdot \hat{T}_{\mathrm{sys, obs}} (t) \\
&= T_{\mathrm{sys}}(\mathrm{1^{st}}) \cdot \left[m\ \hat{T}_{\mathrm{WVR, obs}} (t)+b\right]
\end{split}
\end{equation}
where $T_{\mathrm{sys}}(\mathrm{1^{st}})$ are \Tsys values used to normalize each antenna and each type of observing targets. $m$ and $b$ are the slope and intercept of the fitted linear function. For making the extrapolation, we also sample and average the WVR data every 10 seconds to be consistent with our fitting parameter choice. An example of extrapolated \Tsys for one antenna is shown in Fig. \ref{fig:Tsys_WVR_extrap}. As we can see, the extrapolated continuous \Tsys is consistent with the original discrete \Tsys values for all 4 spectral windows. The trend is also quite continuous with no obvious glitches due to the measurement noise. The trends for all 4 spectral windows are similar. 

Examples of fitting and \Tsys extrapolation for other datasets are shown in Appendix \ref{sec:extra_figure}. We can see for all datasets that \TsysNorm and \TwvrNorm have a tight linear correlation but with different slopes and intercepts. The extrapolation also works well for most of the data sets. 

\subsection{Extrapolate \Tsys with Other WVR Channels and PWV}
\label{sec:wvr_chans}


\begin{table}
	\caption{RMS of the \Tsys residual from the fitting }
	\label{tab:fitting_rms}
	\begin{threeparttable}
		\movetableright=-0.5 in
		\begin{tabular}{ccccc}
			\hline \hline
			\diagbox[width=1.2 in, height=0.4 in]{WVR chans}{\Tsys spws} & 17 & 19 & 21 & 23 \\
			\hline
			0 & 1\% & 1.1\% & 0.9\% & 0.9\% \\
			1 & 0.9\% & 1.1\% & 0.8 \% & 0.8\% \\
			2 & 1\% & 1.1\% & 0.9\% & 0.9\% \\
			3 & 1.4\% & 1.7\% & 1.3\% & 1.3\% \\
			PWV$_{\mathrm{los}}$ & 0.8\% & 0.9\% & 0.7\% & 0.7\% \\
			\hline
		\end{tabular}
		\begin{tablenotes}[flushleft]
		\item The root mean square (RMS) of the residual from the linear fitting using \Twvr from different WVR channels and calculated PWV values along the line of sight (without elevation correction) for dataset Band8. The PWV$_{\mathrm{los}}$ value for dataset Band8 is $\sim$ 1.05 mm. 
		\end{tablenotes}
\end{threeparttable}
\end{table}

The ALMA WVRs have 4 filter channels at frequencies (184.19, 185.25, 186.485 and 188.51 GHz respectively) close to the 183GHz H$_2$O line \citep{Hills_memo352}. These channels have different sensitivities to the line-of-sight water content (PWV) and hence \Tsys, depending on the actual PWV at the observing time. In this section, we explore how the different parameter choices will affect the correlation between \Tsys and \Twvr. In Section \ref{sec:Tsys_Twvr}, we selected \Twvr in channel 1 to track \Tsys. Here we compare how well \Twvr from different WVR channels track the \Tsys from different spectral windows by calculating the scatter of the data residual from the fitting (Table \ref{tab:fitting_rms}). We can see that for these observing conditions (PWV$_{\mathrm{los}}$ of 1.05 mm), the normalized \TwvrNorm from different WVR channels have a similarly tight correlation with normalized \TsysNorm with scatter of $\sim$ 1\%. For dataset Band8, \TwvrNorm from WVR channel 1 gives us the tightest linear correlation. We will discuss the reason later in this section.

For single dish telescopes such as APEX and JCMT, WVR data has been used to continuously track optical depth at the observed frequencies \citep[e.g.][]{Dempsey_2013}. The method converts \Twvr values from multiple WVR channels into a single PWV value and use it to track the optical depth at any given time, which reduces the effect of measurement noise from a single channel. Given what we are doing is similar, as \Tsys is mostly affected by the change in atmospheric optical depth, we can try to use PWV along the line of sight (PWV$_{\mathrm{los}}$) instead of \Twvr from a specific channel to track \Tsys. We calculate the PWV$_{\mathrm{los}}$ by fitting the Lorentz profile for the water line given the \Twvr from multiple WVR channels. We then normalize the PWV$_{\mathrm{los}}$ values the same way as we do for \Twvr (Eq. \ref{eq:Tsys_Twvr_norm}). The scatter of fit residual using PWV$_{\mathrm{los}}$ is also listed in Table \ref{tab:fitting_rms}. As we can see, PWV$_{\mathrm{los}}$ actually gives a tighter correlation, which is consistent with our expectation since it is less affected by the measurement noise from the single channel. 

In our later analysis to apply the continuous \Tsys in data calibration, we select WVR channels to maximize the following weighting function
\begin{equation}
w = \overline{T}_{\mathrm{WVR}} (\overline{T}_{\mathrm{WVR}}-275)
\end{equation}
where $\overline{T}_{\mathrm{WVR}}$ is the averaged \Twvr in one channel and 275 K is the approximate atmosphere temperature. The principle for this selection criterion is to make \Twvr neither too small to be robust against noise (in the case of low opacity) nor too large to be saturated (in the case of high opacity). Based on this criterion, we generally select WVR channel 0 or 1 for datasets in our analysis. 


\subsection{Fewer Atm-cal scans to Fit the Relation}
\label{sec:fewer_Tsys}

\begin{figure}[tbh!]
	\gridline{
		\fig{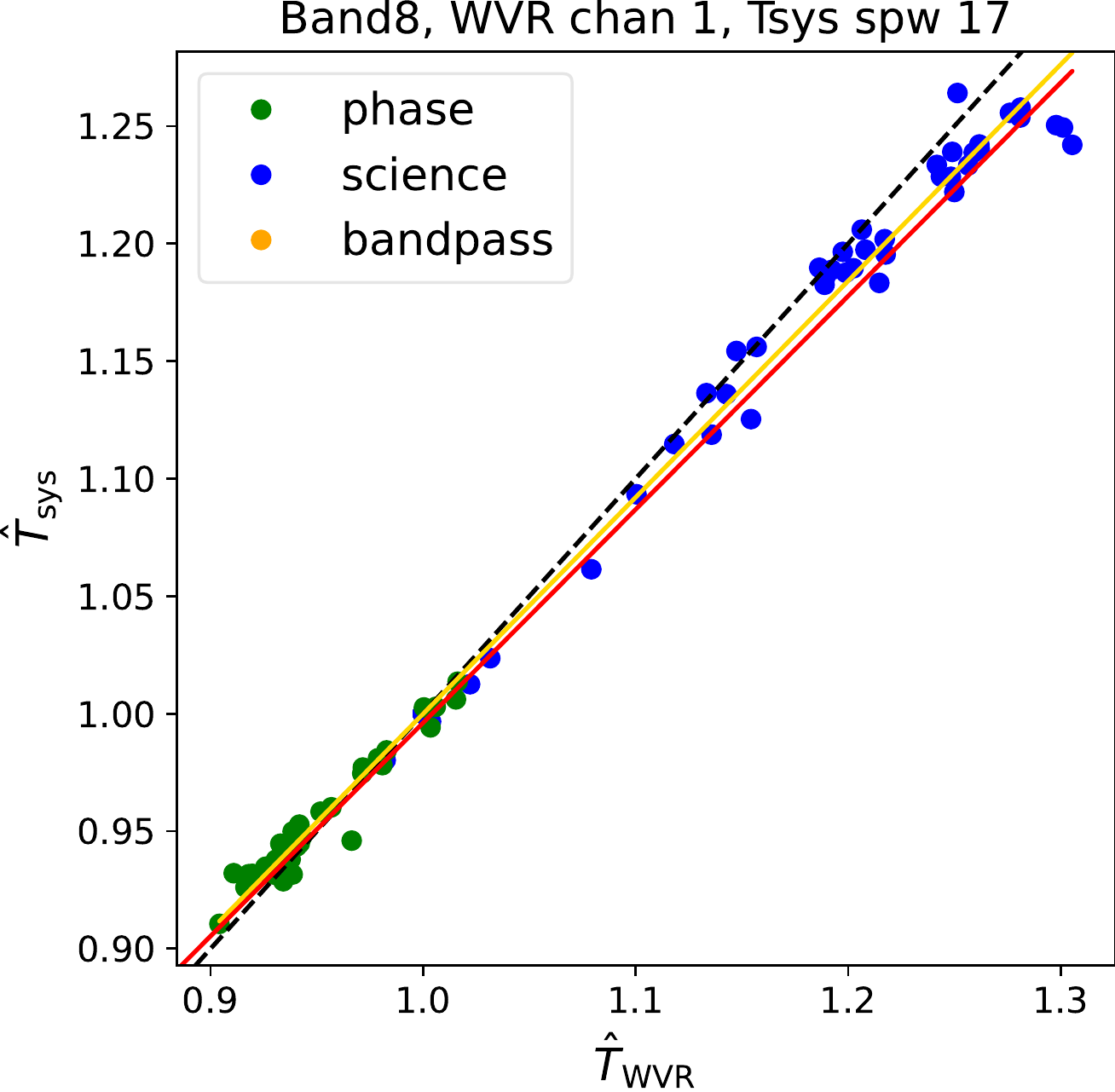}{0.8\linewidth}{}
	} 
	\vspace{-2\baselineskip}
	\caption{The correlation between  \TsysNorm and \TwvrNorm for the 4 Atm-cal scans we selected to fit the linear relation. The red and gold lines are the fitting relation with all Atm-cal scans or just 4 Atm-cal scans. The black dashed line indicates the 1-to-1 relation. We can see the fitting using all Atm-cal scans are almost the same as just using 4 scans. }
	\label{fig: Tsys_WVR_norm_part}
\end{figure}

As discussed in Section \ref{sec:Tsys_Twvr}, the \Tsys in different spectral windows have different linear relations with \Twvr. 
As mentioned, a goal is to reduce the number of \Tsys measurement scans within each observation to increase observing efficiency. However, this leads to less data to fit the relations of \Tsys to \Twvr or PWV. In Section \ref{sec:highFreq} we investigate determining the relations from atmosphere opacity models, but here we test the reliability of fitting the relations to a small number of \Tsys measurement scans. 
Since we need to calculate the normalized \Tsys, we need at least two Atm-cal scans to fit the linear correlation. To make the fitting more robust, we use 4 Atm-cal scans for the fitting with 2 from phase target and 2 from the science target. The 4 Atm-cal scans give us 2 independent \TsysNorm values if we normalize the \Tsys from phase and science target independently. For the Atm-cal scan selection, we select 2 Atm-cal scans at the start and 2 Atm-cal scans in the middle. One example of the fits using 4 Atm-cal scans is shown in Fig. \ref{fig: Tsys_WVR_norm_part}. As we can see,  the fits based on data from all Atm-cal scans have almost no difference from the fits based on only 4 Atm-cal scans. The scatter of all the data points around the new relation has almost the same scatter of 1\%. 

\begin{figure*}[tbh!]
	\gridline{
		\fig{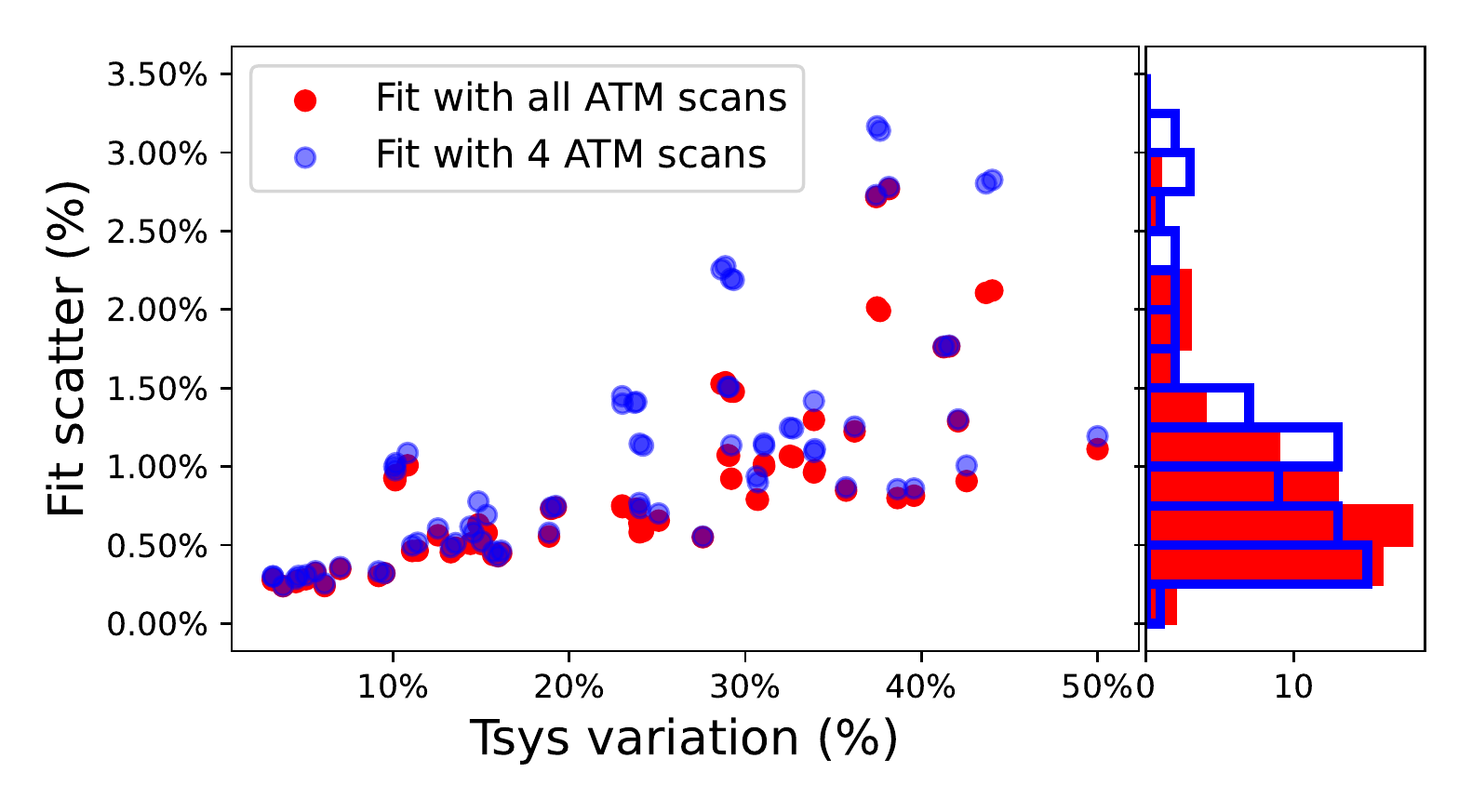}{0.8\linewidth}{}
	} 
	\vspace{-2\baselineskip}
	\caption{The scatter of data points around the \Tsys vs \Twvr fits with all the antennas versus the maximal difference in $\hat{T}_{\mathrm{sys}}$ values for each \Tsys spw of each dataset. The red and blue points are from fitting with all Atm-cal scans or just 4 Atm-cal scans respectively. The histogram at the right side shows the distribution of the fit scatters using the two different methods. We can see the scatter of the fitting only increases slightly using just 4 Atm-cal scans. }
	\label{fig:fit_statistics}
\end{figure*}

Fits using 4 Atm-cal scans for other data sets are also shown in Appendix \ref{sec:extra_figure}. We can see that the fits do not change much for almost all the datasets except Band9b1, which we will in Section \ref{sec:highFreq}. In Fig. \ref{fig:fit_statistics}, we plot the relative scatter around the fit versus the maximal difference divided by mean value of the \Tsys for every \Tsys spw of all the data sets. As we can see, the scatter using both fitting methods is generally below 3\%. Fitting with just 4 scans only slightly increases the scatter compared with fitting with all scans. From this quantitative comparison, we can see it is viable to reduce the number of discrete \Tsys measurements when using \Twvr to track \Tsys. 

\subsection{Normalize Only to the Science target}
\label{subsec:sci_norm}

\begin{figure}
	\gridline{
		\fig{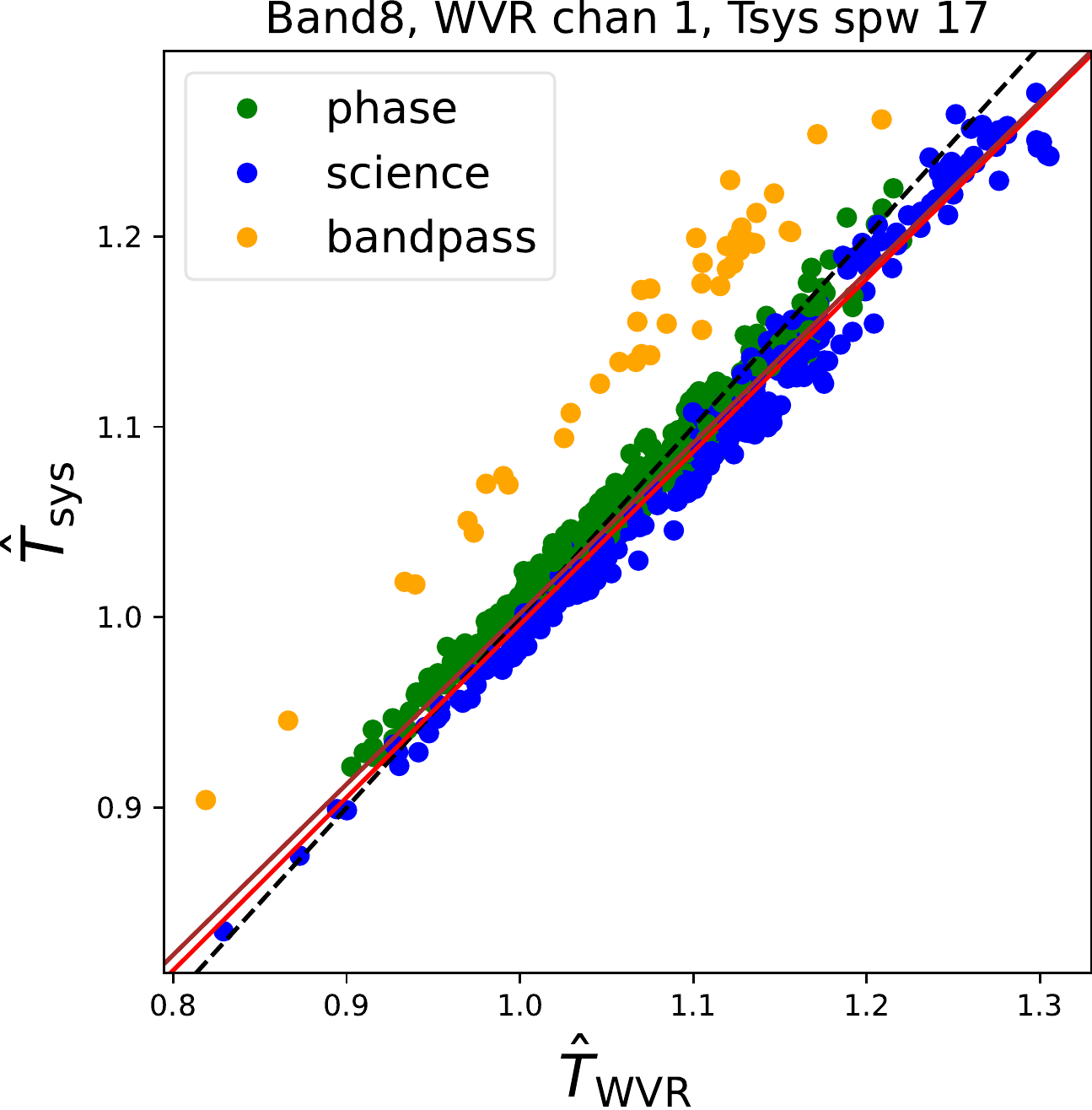}{0.9\linewidth}{}	
	} 
	\vspace{-1.5 \baselineskip}
	\caption{\Tsys versus \Twvr normalized to the first science Atm-cal scan for the dataset Band8. The black dashed line is the 1-to-1 relation. The brown solid line is the linear fitting to the data excluding the bandpass data. The red solid line is the original fitting relation to the data normalized to each type of observing target.  We can see the two fitting relation are almost the same. The bandpass data for spw 0 has a significant offset from the fitted relation, which is due to the elevation difference between bandpass target and phase-cal/science targets (see discussion in Section \ref{subsec: Tsys_Twvr_ATM}). }
	\label{fig:Tsys_WVR_norm_sci}
\end{figure}


For some ALMA data, \Tsys for the phase-cal target is not measured. Instead, the calibration uses the nearest science \Tsys values as the phase-cal \Tsys. If we can normalize all \Tsys values to the first \Tsys of the science target instead of the first \Tsys of each type of observing target itself, we can further reduce the number of \Tsys measurements and thus no longer need to measure \Tsys for phase-cal with our new method.

In this case, the normalized \Tsys is calculated as 
\begin{equation}
\label{eq:Tsys_Twvr_norm_sci}
\begin{split}
\hat{T}_{\mathrm{sys, obs}} (t) &= \frac{T_{\mathrm{sys, obs}}(t)}{T_{\mathrm{sys, sci}}(1^{\mathrm{st}})} \\
\hat{T}_{\mathrm{WVR, obs}} (t) &= \frac{T_{\mathrm{WVR, obs}}(t)}{T_{\mathrm{WVR, sci}}(1^{\mathrm{st}})}
\end{split}
\end{equation}
where $\hat{T}_{\mathrm{sys, obs}} (t)$ is the normalized \Tsys averaged along the spectral axis and $\hat{T}_{\mathrm{WVR, obs}} (t)$ is the normalized \Twvr. An example of \TsysNorm versus \TwvrNorm using the new normalization method are shown in Fig. \ref{fig:Tsys_WVR_norm_sci}. We can see that phase-cal and science target generally follows the same linear trend, which is consistent with our expectation since phase-cal and science targets are close in elevation. In contrast, we see offsets between the trends of the bandpass target and phase-cal/science targets. We also expect this to happen since bandpass target usually has significant different elevations from the phase-cal/science targets. We will further discuss the cause of the offsets with the help of atmospheric modeling in Section \ref{subsec: Tsys_Twvr_ATM}. In general, these tests show that we can further reduce the phase-cal and bandpass \Tsys measurements as we can derive it from \Tsys measurements for only the science target. 

Note that for early ALMA cycles, the \Tsys measurements are purely done for the phase-cal target. The \Tsys for the science target is then assumed to be the same as the \Tsys for the closest phase-cal scan. As we can see from this section, even though the phase-cal and science targets have different elevations, they generally follow the same  \TsysNorm vs \TwvrNorm linear relation. Therefore, we can better extrapolate the science \Tsys from the phase-cal \Tsys using the fitted linear relation. 

\subsection{Test with significant opacity and large \Tsys variation}
\label{sec:highFreq}

\begin{figure*}
	\gridline{
		\fig{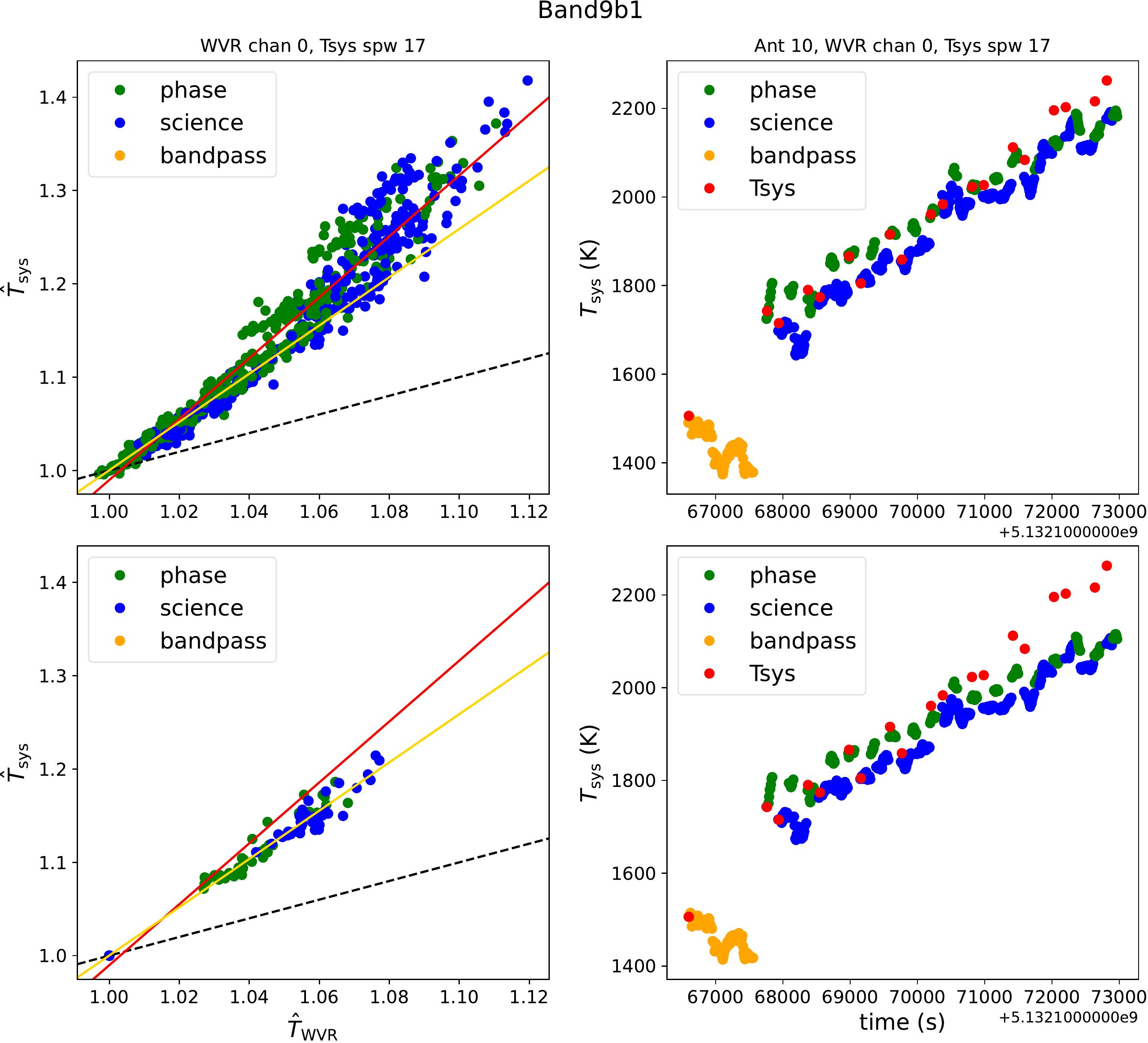}{0.8\linewidth}{}
	} 
	\vspace{-2\baselineskip}
	\caption{(Left) \TsysNorm (spw 17) versus \TwvrNorm (channel 0) for dataset Band9b1. The upper left panel shows \TsysNorm versus \TwvrNorm in spw 17 using all Atm-cal scans while the lower left panel shows the correlation with selected 4 Atm-cal scans. The red and yellow lines are the fits derived using all Atm-cal scans and just 4 Atm-cal scans. (Right) The measured and extrapolated \Tsys for different observing targets as a function of time. Red points are measured \Tsys values. \Tsys in upper panel is extrapolated based on fits with all Atm-cal scans while \Tsys in lower panel is extrapolated based on fits using only selected 4 Atm-cal scans. We can see in this case we will underestimate the \Tsys value if we just use part of Atm-cal scans to fit the correlation. }
	\label{fig:Tsys_WVR_highfreq}
\end{figure*}

We would expect the linear relation between \Tsys and \Twvr holds when $\tau_{\mathrm{sky}}$ is small. In this case, we would have
\begin{equation}
\label{eq:Tsys_lowtau}
T_{\mathrm{sys}} \approx T_{\mathrm{rx}} + T_{\mathrm{sky}} = T_{\mathrm{rx}} + C \times  T_{\mathrm{WVR}} + T_{\mathrm{sky, dry}}
\end{equation}
where $C$ is a constant. However, at higher frequencies such as Band 9 and 10, $\tau_{\mathrm{sky}}$ are quite high and we can no longer ignore the opacity term in \Tsys (e.g. Eq. \ref{eq:Tsys_simplified}). In this case, the increase in \Tsys is dominated by the increase in $\tau_{\mathrm{sky}}$ in the exponential form. 

We test if the \TsysNorm vs \TwvrNorm linear relation still holds on dataset Band9b, which has large $\tau_{\mathrm{sky}}$ and \Tsys range ($\sim$ 50\%). In Fig. \ref{fig:Tsys_WVR_highfreq}, we show \Tsys fitting and extrapolation using all Atm-cal scans or just 4 Atm-cal scans for one measurement set in this project. We can clearly see there is a difference in the fitting functions derived from all Atm-cal scans or just 4 Atm-cal scans. It seems the slope becomes steeper due to data points with higher \Tsys values, which are not included if we use just 4 scans. This is also reflected in the extrapolation plot at the right side of Fig. \ref{fig:Tsys_WVR_highfreq}, as the predicted \Tsys is lower than the measured \Tsys for higher \Tsys values. 

This is consistent with our expectation that the slope of \Tsys vs \Twvr relation is increasing. We would expect the curving-up feature also happens to other datasets but we do not have large enough \Tsys ranges in the other datasets we analyzed.

\section{Atmospheric Transmission at Microwave (ATM) Modeling}
\label{sec:ATM_modeling}

In the previous sections, we fit \Tsys vs \Twvr heuristically and use the fitted relation to extrapolate the \Tsys continuously. We find in most cases the correlation between \Tsys and \Twvr is linear. However, the exact slopes and intercepts of the correlations vary across different frequencies. Although we can use less than 4 Atm-cal scans to fit the relation for each spw in each dataset, there might be cases when our selected Atm-cal scans have similar \Tsys values, and hence might give us inaccurate fitting relation among small \Tsys ranges. Furthermore, as described in Section \ref{sec:highFreq}, the linear approximation becomes insufficient when the opacity becomes significant or \Tsys variations become large. A more robust method is to use an atmospheric opacity modeling code and an estimate of the \Trx and other static contributions to predict the \Tsys vs \Twvr relation at the relevant frequencies, elevation and PWV ranges of the observation. In this section we use the Atmospheric Transmission at Microwave frequencies (ATM) model \citep{Pardo_2001} to predict the \Tsys vs \Twvr relation for various datasets and compare the results with our heuristic method. 

\subsection{Modeling \Tsys spectrum}
\label{subsec:Tsys_spectrum}

\begin{figure*}
	\gridline{
		\fig{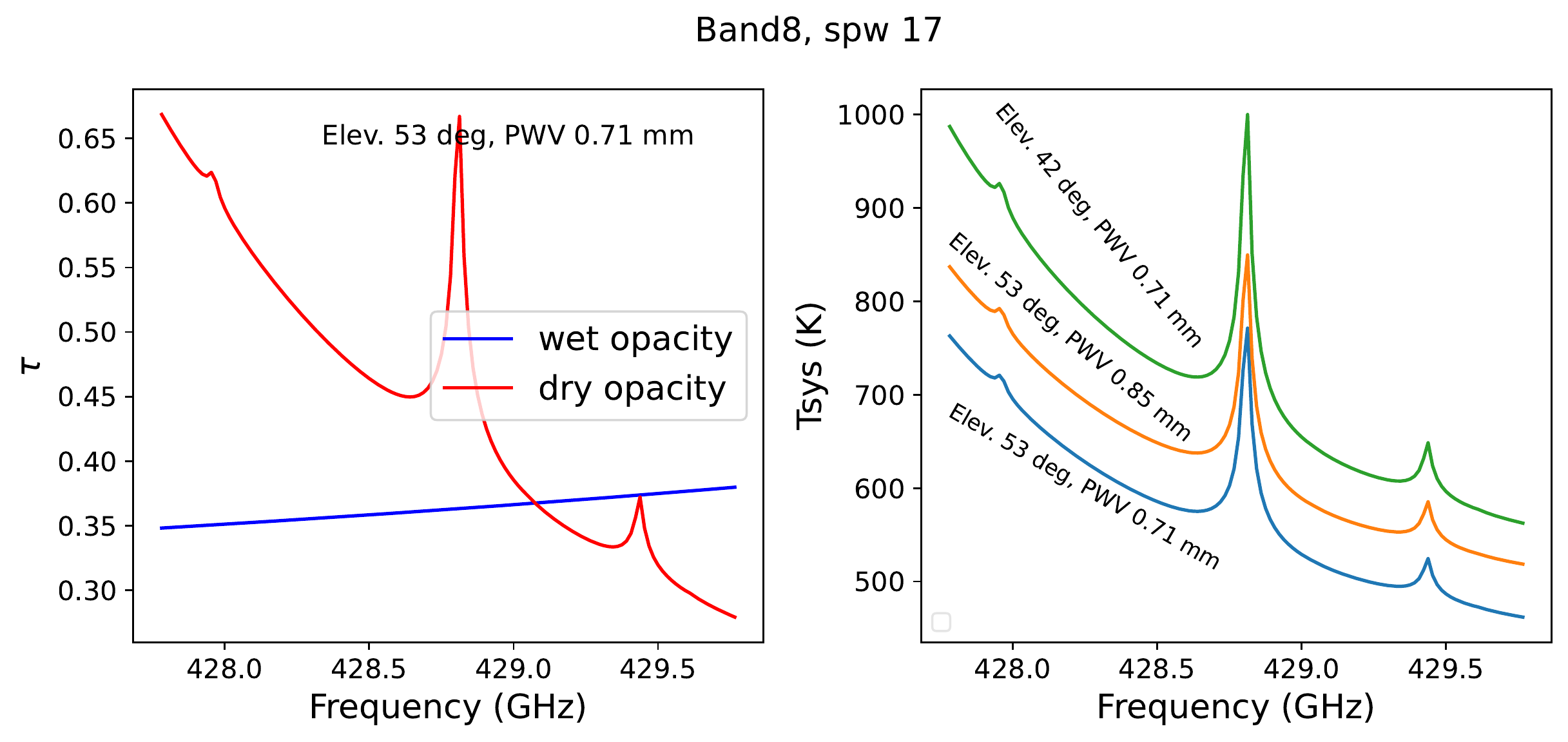}{0.8\linewidth}{}}
	\vspace{-3\baselineskip}
	\gridline{
		\fig{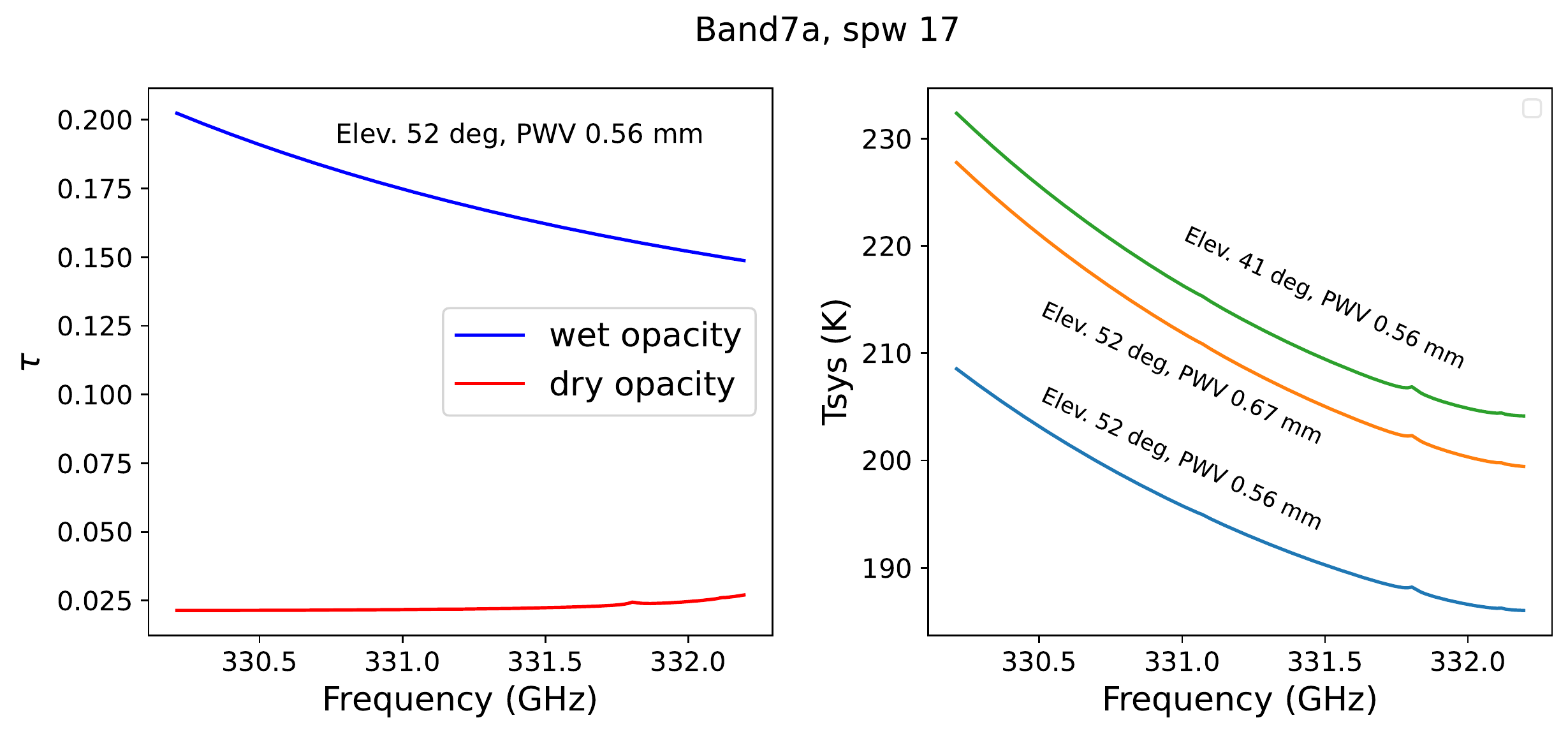}{0.8\linewidth}{}}
	\vspace{-2\baselineskip}
	\caption{The ATM modeling opacity $\tau_{\mathrm{sky}}$ and \Tsys spectrum for dataset Band8 (upper) and Band7a (lower). For each row, the left panel shows the $\tau_{\mathrm{sky}}$ spectrum for the wet and dry component. We can see that the dry component is significant for dataset Band8 but the water component is dominant for Band7a. The right panel shows the modeled \Tsys spectrum for 3 sets of different PWV and elevation values. The blue line is the baseline while we increase PWV or airmass value by 1.2 respectively for orange and green lines. We can see increasing airmass will increase \Tsys faster since it increase \Tsky from both wet and dry components.  }
	\label{fig:ATM_spectrum}
\end{figure*}

We note that in previous sections when we explore the correlation between \Tsys and \Twvr, we average \Tsys for each spectral window along its spectral axis. Therefore, when we extrapolate the continuous \Tsys, we are assuming that the \Tsys spectrum does not vary significantly. In this subsection, we test this assumption with ATM modeling on the two representative datasets we have, Band8 and Band7a. We use the version of ATM included in CASA \citep{McMullin_2007, Emonts_2020, Bean_2022}, accessed via a helper function \texttt{plotAtmosphere}\footnote{\url{https://safe.nrao.edu/wiki/bin/view/ALMA/PlotAtmosphere}} to generate \Tsys and \Twvr spectra for the frequency ranges of a given spw in the data. We set most of the parameters to the default for the ALMA site (height 5000 m, pressure 557 mb and temperature 274 K). For each dataset, we set the PWV and elevation values to be the same as the value of the first Atm-cal scan for the science target as our start point. Note that \texttt{aU.plotAtmosphere} only gives \Tsky and $\tau_{\mathrm{sky}}$. Therefore, we calculate the \Tsys spectrum from the modeled \Tsky and $\tau_{\mathrm{sky}}$ using Eq. \ref{eq:Tsys_def} by assuming \Trx to be 100 K. 

We show our modeled results in Fig. \ref{fig:ATM_spectrum}. As we can see, spw 17 for dataset Band8 has a significant dry opacity contribution as it sits at one of the O$_2$ lines. On the contrary, dataset Band7a is dominated by the wet component. We then increase the PWV and airmass ($1/\sin{m_{\mathrm{el}}}$, $m_{\mathrm{el}}$ is the elevation) by a factor of 1.2 to see if they have different effects on increasing the \Tsys spectrum. For both dataset Band8 and Band7a, we can see increasing airmass is more effective in increasing the overall values of the \Tsys spectrum. This is what we expect since increasing airmass will increase both wet and dry opacity while increasing PWV only increases the wet opacity. If we compare dataset Band8 with Band7a, we can see the difference between increasing PWV and airmass is more significant for Band8 as the spectral window has significant dry opacity contribution. 

The change of \Tsys spectrum shape is generally small by increasing the PWV or airmass by 20\%. However, a small change is noticeable when increasing airmass for spw 17 in dataset Band8, due primarily to the significant $\tau_{\mathrm{sky}}$ from the dry component and its large variation across the spw. Not tracking such small \Tsys spectrum shape changes will have negligible impact on continuum observations, and for spectral lines the error in the extrapolation based on \Twvr or PWV will be within $\sim \pm$ 3\% (see Section \ref{subsec:additional} for more discussion). A future improvement might be to correct the data spectrally rather than using a single channel-averaged value per timestamp.


\subsection{Reproduce the Observed \TsysNorm vs \TwvrNorm correlation}
\label{subsec: Tsys_Twvr_ATM}
\begin{figure}
	\gridline{
		\fig{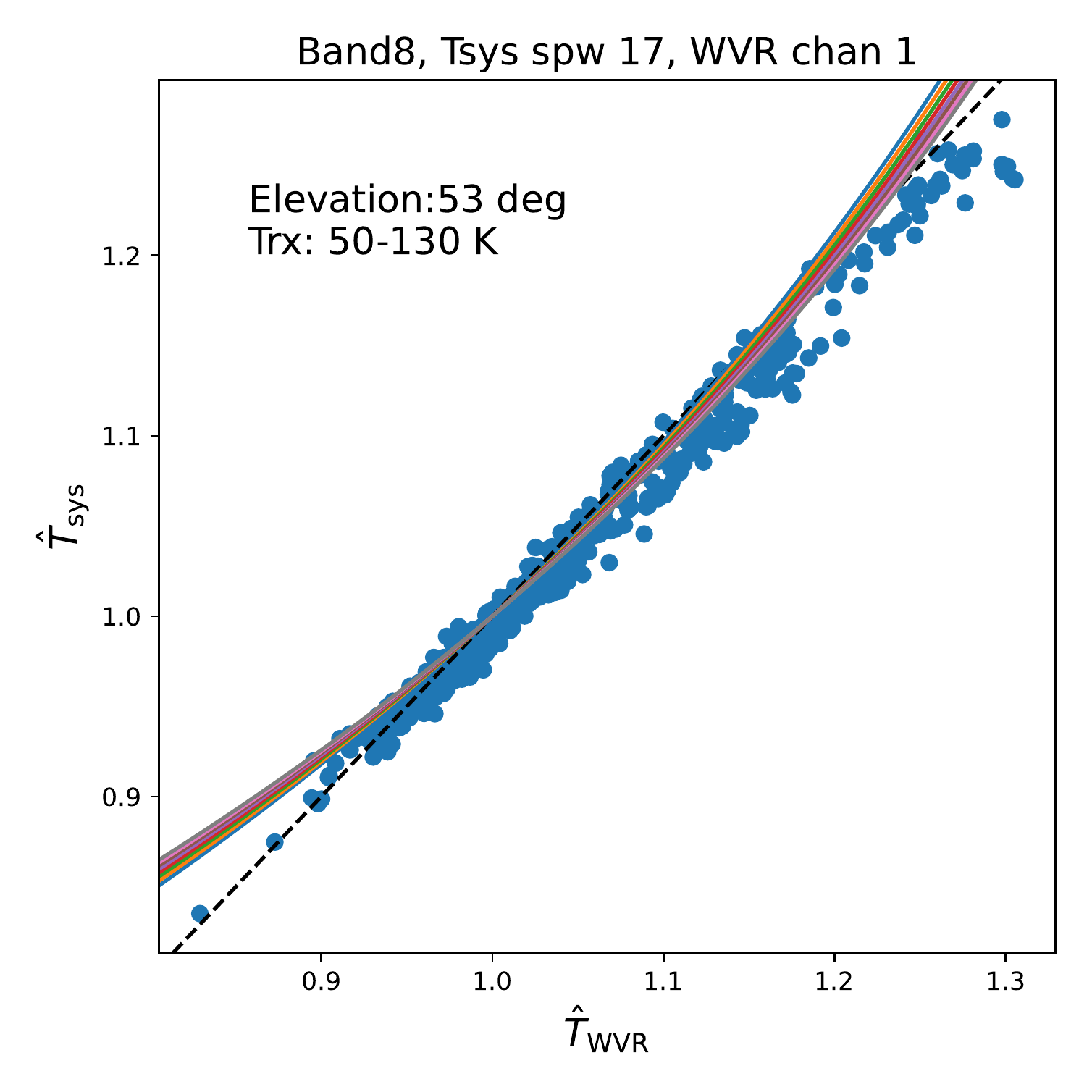}{0.9\linewidth}{}}
	\vspace{-3\baselineskip}
	\caption{
		The correlation between \TsysNorm and \TwvrNorm (Fig. \ref{fig: Tsys_WVR_norm}) overlaid by the predicted correlation curves from ATM modeling. Curves of different colors represent modeling using different \Trx values. The modeled \Tsys and \Twvr are normalized to the value when PWV is 0.71 mm, which is the PWV value for the first Atm-cal scan for the science target.  The \Trx range we use for ATM modeling is similar to the \Trx range of dataset Band8. We can see that varying \Trx generally does not affect the correlation we get.  }
	\label{fig:Tsys_Twvr_ATM_Trx}
\end{figure}

In this section we will try to reproduce the observed \TsysNorm vs \TwvrNorm correlation in various cases. According to Eq. \ref{eq:Tsys_simplified}, \Tsys are determined by \Trx and \Tsky. \Tsky can be further determined by the measurement of PWV and elevation through ATM modeling. Therefore, we can generate modeled \Tsys spectrum with given \Trx, PWV and elevation at a certain frequency range. To reproduce the \Tsys measured in observation, we set the frequency range to be the same as the \Tsys spectral window we want to model, with total bandwidth of $\sim$ 2 GHz. The modeled \Tsys spectrum is then averaged to a single \Tsys value as we did with the observations. We also use a similar method to generate \Twvr at different WVR channels with given PWV and elevations (\Twvr is just \Tsky at the WVR channel frequencies). 

For most of the ALMA data, \Trx stays relatively constant throughout the observations. However, different antennas generally have different \Trx values, which might give us slightly different shapes of correlation. Therefore, we first test how varying \Trx could affect the shape of the \TsysNorm vs \TwvrNorm correlation for dataset Band8 spw 17 (Fig. \ref{fig:Tsys_Twvr_ATM_Trx}). The \Tsys and \Twvr are generated with varying \Trx and PWV values but with fixed elevation of 53 deg. We note that dataset Band8 has relatively constant elevations (see Table \ref{tab:data_summary}) for the science target throughout the entire observation, hence the \Tsys variations across time are mostly due the the change in PWV values. After generating the modeled \Tsys and \Twvr, we then normalize both quantities to the values when the PWV value is equal to that of the first science Atm-cal scan for each \Trx value. As we can see in Fig. \ref{fig:Tsys_Twvr_ATM_Trx}, the ATM modeling shows a slight non-linear curvature for a high range of \Tsys, which depends slightly on the assumed \Trx and the elevation. However, the modeling curve is generally within the range of the data scatter. Varying \Trx also gives a similar correlation within data scatter of $\sim$ 1\%. Therefore, \Trx values do not seem to affect the \TsysNorm vs \TwvrNorm correlation we get. 

\begin{figure*}
	\gridline{
		\fig{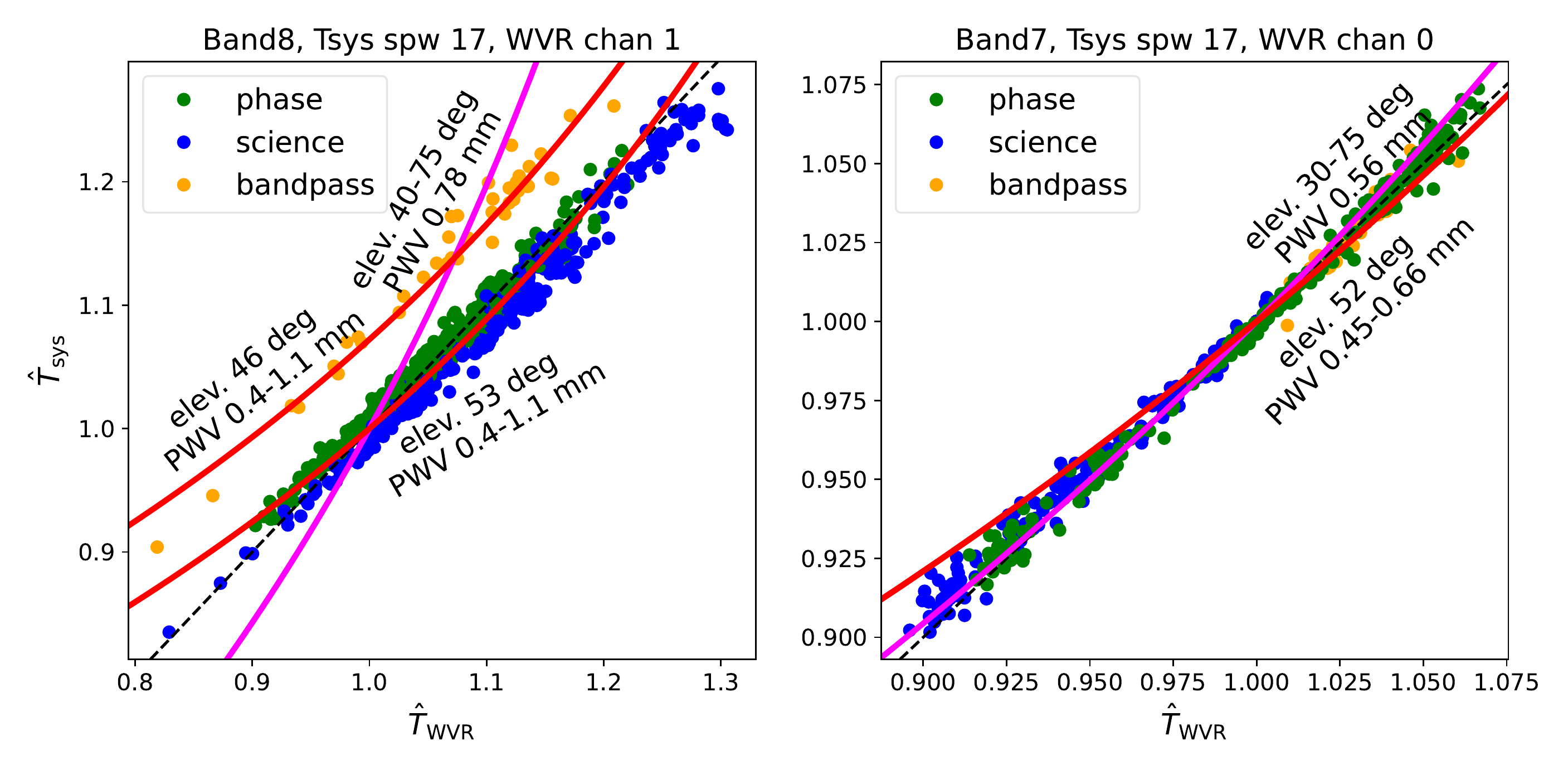}{0.9\linewidth}{}}
	\vspace{-2\baselineskip}
	\caption{The correlation between \TsysNorm and \TwvrNorm overlaid by the predicted correlation curves from ATM modeling for dataset Band8 (left) and Band7a (right). Note that in this case both \Tsys and \Twvr are normalized to the first Atm-cal scan for the science targets (Eq. \ref{eq:Tsys_Twvr_norm_sci}). The magenta line represent varying elevations while keeping the PWV value to be the same and the red line keeps elevation constant but varies PWV. The \Trx are set to be equal to that of the first antenna of the data. The ATM modeled \Tsys and \Twvr are normalized to the value when PWV and elevation are equal to those of the first science Atm-cal scans in the data. We can see that for dataset Band8 where the dry component is significant, varying PWV or elevation gives us correlations of different slopes. Since the bandpass and science/phase targets have different elevations, we therefore see the offsets between \Tsys for these two targets. In contrast, for dataset Band7a where the wet component is dominant, varying PWV or elevation gives us similar \TsysNorm vs \TwvrNorm correlations. In this case, all 3 targets follow the same linear trend. }
	\label{fig:Tsys_Twvr_ATM_Band78}
\end{figure*}

We then explore how varying PWV or elevation can affect the \TsysNorm vs \TwvrNorm correlation. As we have mentioned in Section \ref{subsec:Tsys_spectrum}, varying PWV or elevation might have different effects on changing \Tsys values depending on how significant the dry opacities are at given frequency. Dataset Band8 and Band7a represent two cases where one has significant dry opacity contribution while the other is dominated by the wet opacity. Therefore, it is natural for us to explore what drives the \TsysNorm vs \TwvrNorm correlation in these two cases. Fig. \ref{fig:Tsys_Twvr_ATM_Band78} shows the comparison between modeling and observation for these two datasets where red and magenta lines represent changing PWV and elevation respectively. In this comparison, the measured \Tsys and \Twvr are normalized to the values in the first science Atm-cal scan instead of the first Atm-cal scan for each target (see Section \ref{subsec:sci_norm} for more description). The modeled \Tsys and \Twvr are then normalized to the values when PWV and elevation are equal to those of the first science Atm-cal scan. We can see in both cases the ATM modeling successfully reproduces the observed \TsysNorm vs \TwvrNorm correlation. For dataset Band8, the correlation is mainly driven by varying PWV values as the elevation for the science target stays relatively constant. By varying the elevation, we see a steeper slope of the correlation between \TsysNorm and \TwvrNorm. This is due to the fact that changing elevations (and therefore airmass) will be more effective to change \Tsys values when the dry opacity contribution is significant (see discussion in Section \ref{subsec:Tsys_spectrum}). \Tsys for the bandpass target shows significant offsets from the main trend of phase/science target mainly due to the elevation difference, and hence sits at the red track of a different constant elevation value. Since the single bandpass \Tsys measurement for each antennae has the same elevation but might point towards slightly different part of skys with different PWV values, we see the bandpass data points still follow the track of constant elevation. On the other hand, we get a similar \TsysNorm vs \TwvrNorm correlation by varying PWV or elevation for dataset Band7. This is probably due to the fact that the wet component is dominant at this spectral window, hence changing PWV or elevation achieves a similar effect. We can also see in this case the \Tsys for the bandpass target lie along the same trend as the phase-cal/science targets, which is what we expect since there is no specific parameter variation that could bring these data points out of the linear track. 

\begin{figure}
	\gridline{
		\fig{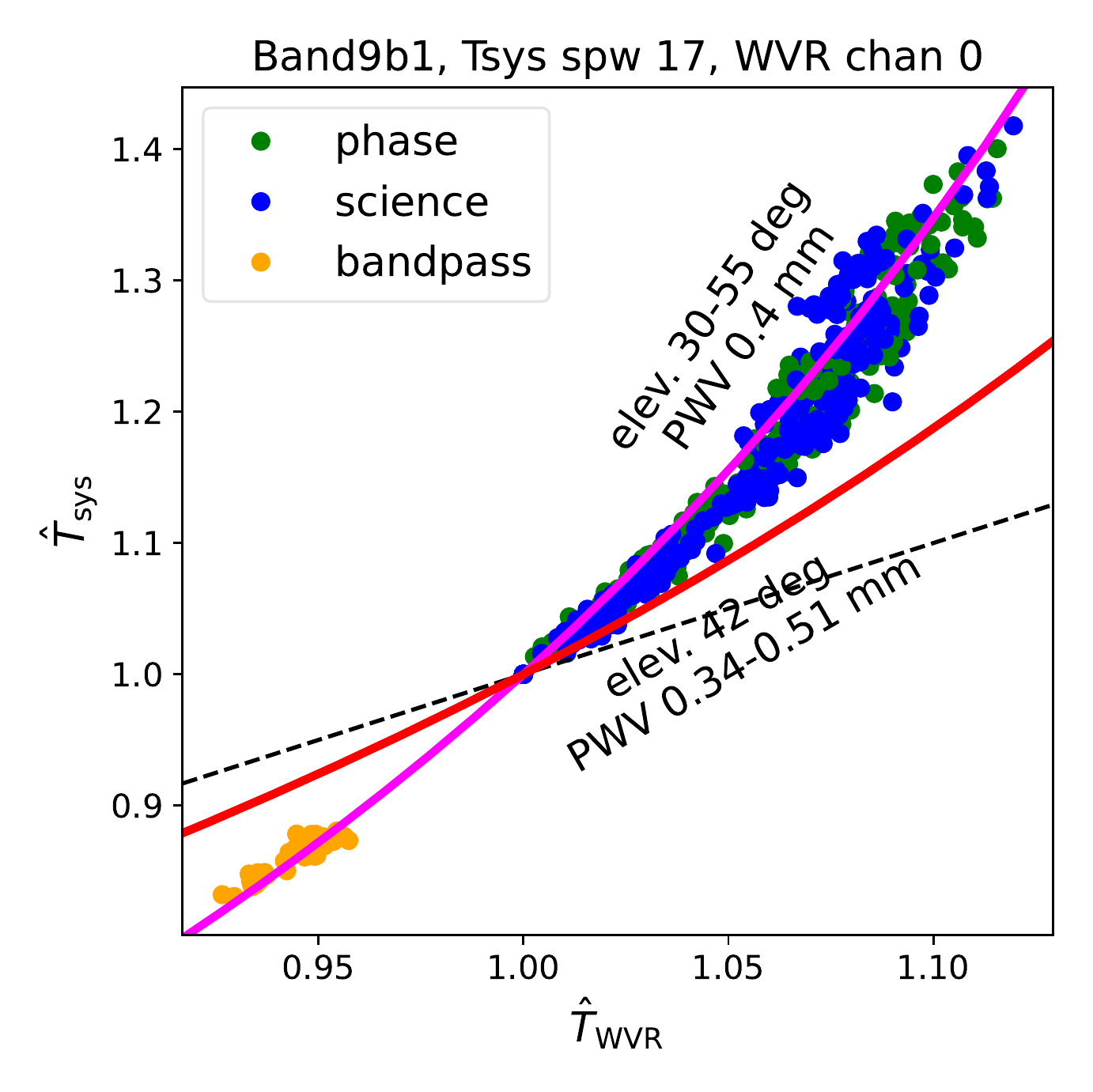}{0.9\linewidth}{}}
	\vspace{-3\baselineskip}
	\caption{Similar plot as Fig. \ref{fig:Tsys_Twvr_ATM_Band78} but for dataset Band9b1. The dry component is also significant for this dataset. We can see that Band9b1 has a relatively constant PWV while the \Tsys and \Twvr variation is mainly due to changing elevations.  }
	\label{fig:Tsys_Twvr_ATM_Band9b1}
\end{figure}

As we have discussed in Section \ref{sec:highFreq}, dataset Band9b1 shows a non-linear \TsysNorm vs \TwvrNorm correlation as the \Tsys variation becomes significantly large ($\sim$ 40\%). Therefore, it is also worth testing if we can reproduce the curving feature for this dataset. We show the comparison in Fig. \ref{fig:Tsys_Twvr_ATM_Band9b1} using the same method described in the previous paragraph. As we can see, this dataset also has significant contribution from the dry opacity and hence shows different slopes when varying PWV or the elevation values. The observed \Tsys generally agrees well with the ATM modeling relation with fixed PWV values. This suggests the \Tsys variation shown for this dataset is mainly due to the elevation change. This is consistent with what we see in Fig. \ref{fig:Tsys_WVR_highfreq} as \Tsys is smoothly increasing as the function of time without any short-time fluctuation. The bandpass data points also lie along the fixed PWV trend but with smaller values, which is probably due to the larger elevation of the bandpass target. We also see some second-order scatter around the fixed PWV line for higher \Tsys values, which might be due to the intrinsic scatter of PWV values during the observation. However, to first order we can just measure \Tsys once and predict the following \Tsys based on the elevation change during the observation. 

\subsection{General Applicability of the current method and future direction for improvement}
\begin{figure*}[tbh!]
	\gridline{
		\fig{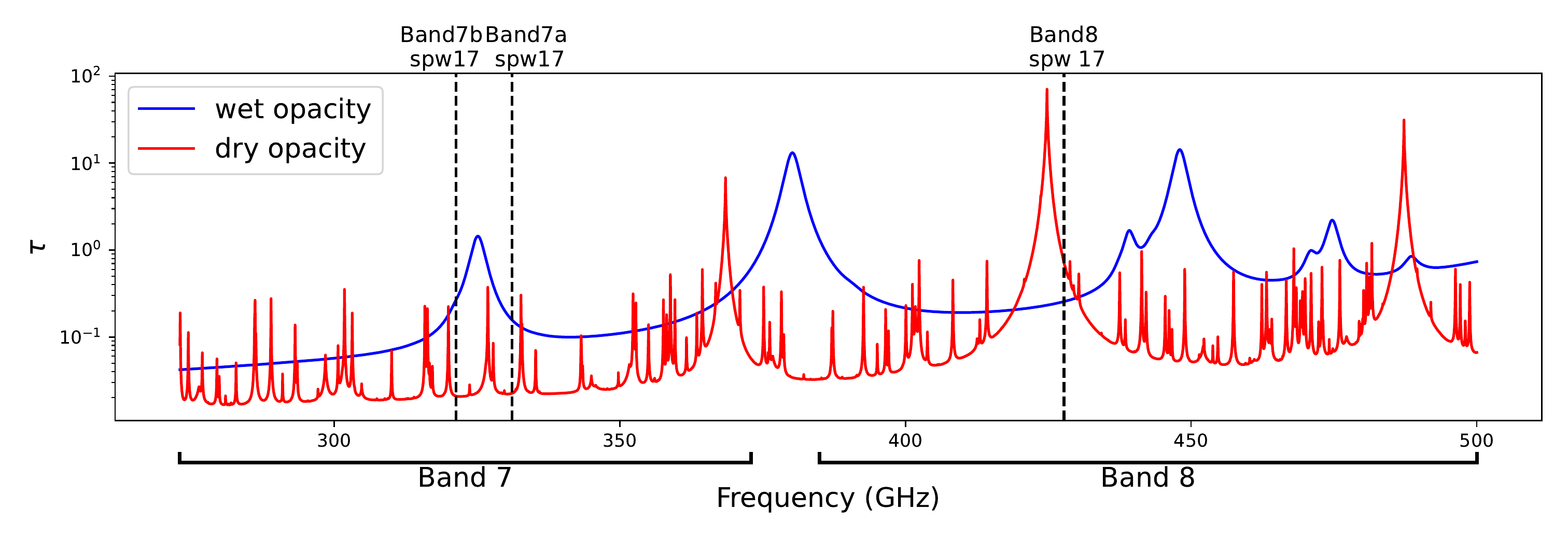}{\linewidth}{}
	} 
	\vspace{-4\baselineskip}
	\gridline{
	\fig{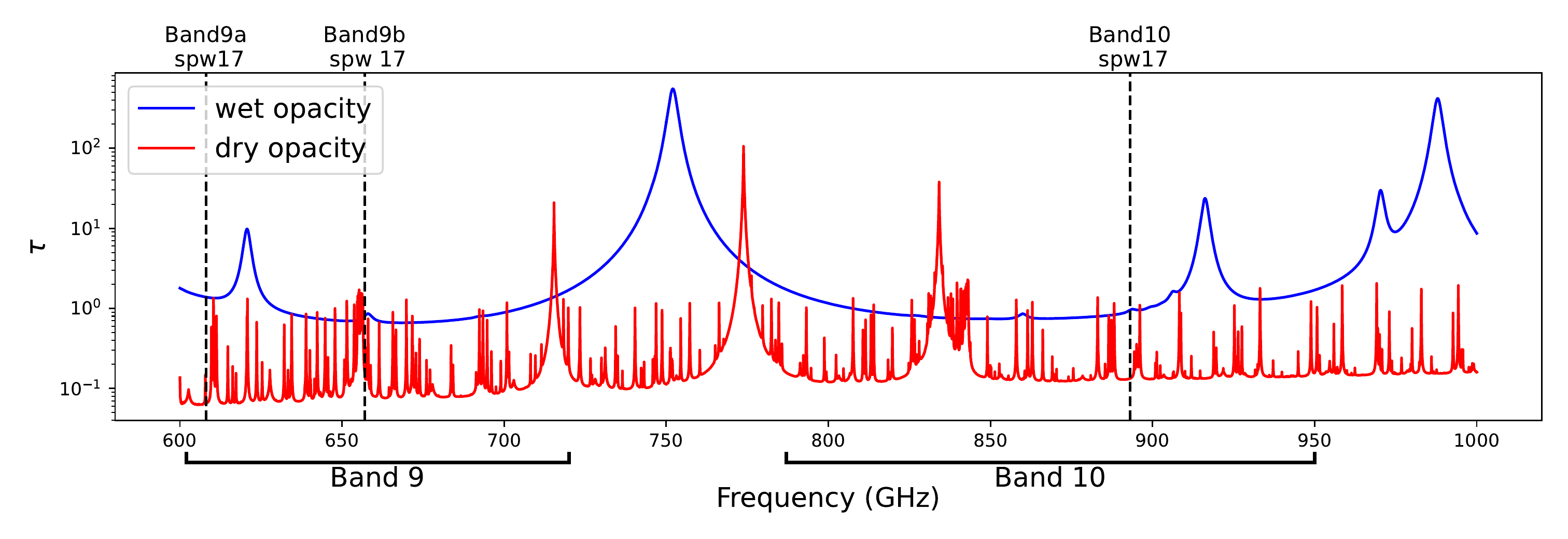}{\linewidth}{}
} 
	\vspace{-2\baselineskip}
	\caption{The wet and dry opacity from ATM modeling for the entire ALMA high frequency bands (from Band 7 to 10). We set the PWV value to be 0.5 mm and elevation to be 50 deg. The dashed line indicates the position of spectral window of ALMA data we have. We can see all of our dataset except for Band8 and Band9b are dominated by the wet component, which means changing PWV or elevations should give the similar \TsysNorm vs \TwvrNorm correlation for these datasets.  }
	\label{fig:tau_modeling}
\end{figure*} 

In the previous section, we have successfully reproduced the \TsysNorm vs \TwvrNorm correlation for 3 datasets with ATM modeling. However, we also find the correlation is not driven by a single parameter. For both dataset Band8 and Band9b1 where dry opacity is significant, the correlation is driven either by varying PWV or elevation. If both PWV and elevation have significant variation, we would not be able to get the tight correlation since the data points are driven up and down along different tracks. In contrast, for dataset Band7a where wet opacity is dominant, we can expect a tight \TsysNorm vs \TwvrNorm correlation even though both PWV and elevation have significant variation, as they are moving data up and down along the similar track. Therefore, it is safer to apply our current heuristic method to datasets observed at frequencies where wet opacity is dominant. In Fig. \ref{fig:tau_modeling}, we model the wet and dry opacity spectrum covering the entire ALMA high frequency bands from Band 7 to Band 10. We assume PWV of 0.5 mm, which is a typical value for high-frequency ALMA observations, and elevation of 50 deg. For the wet opacity, we see several smooth line features, which indicate the presence of the H$_2$O line. For the dry opacity, we generally see a lot of Ozone lines as narrow spikes. These Ozone lines are generally much narrower than the typical bandwidth of a spw of 2GHz and hence have a relatively small effect on the averaged \Tsys values. On the other hand, we also see some broader spikes caused by O$_2$ lines. One of our datasets, Band8, sits right at the wings of one O$_2$ line and thus has significant dry opacity contribution. For all of our datasets used in this paper, only datasets Band8 and Band9b show significant dry opacity contribution, which is consistent with our expectation as all of the other datasets in this paper exhibit a tight linear correlation (Fig. \ref{fig:Tsys_WVR_spw17_ant10_summary}, \ref{fig:Tsys_WVR_spw17_ant10_band9} and \ref{fig:Tsys_WVR_spw17_ant10_single}). 

However, we can expect that in lots of cases, the \TsysNorm vs \TwvrNorm might not follow a tight linear correlation due to various reasons mentioned above. In these cases, the best way is to directly derive \Tsys from the ATM modeling. As shown in Section \ref{subsec: Tsys_Twvr_ATM}, with known \Trx, PWV and elevations for each Atm-cal scan, we can successfully reproduce the \Tsys measured in observations. \textit{In the future, the best strategy for tracking \Tsys is to derive PWV values from continuous \Twvr measurements at different WVR channels. By combining PWV, elevation and \Trx values, we can then reproduce the continuous \Tsys throughout the observation.}

\section{AC \& SQLD Data in Tracking \Tsys}
\label{sec:AC}

\begin{figure*}[tbh!]
	\gridline{
		\fig{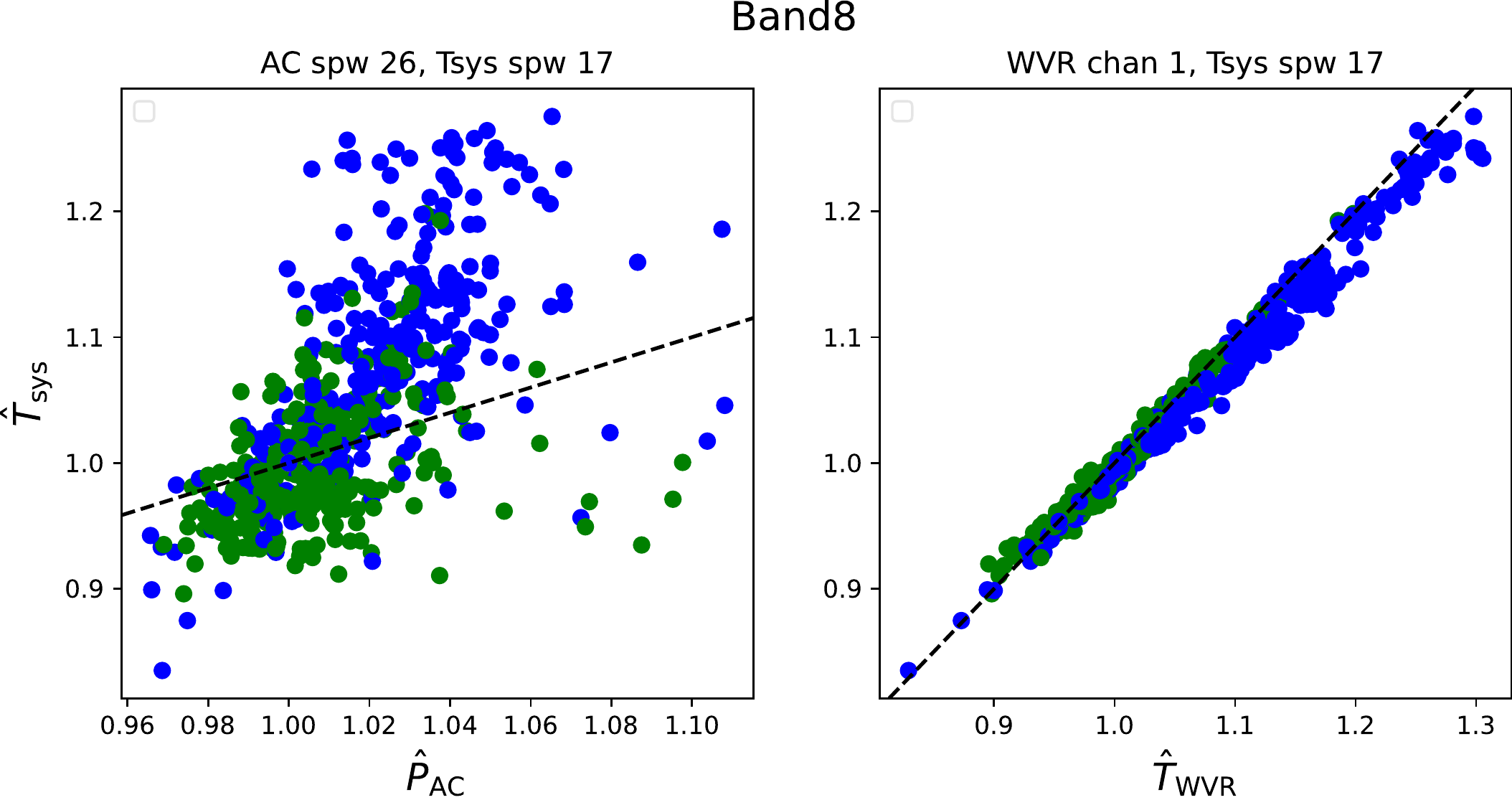}{0.9\linewidth}{}
	} 
	\vspace{-3\baselineskip}
	\gridline{
	\fig{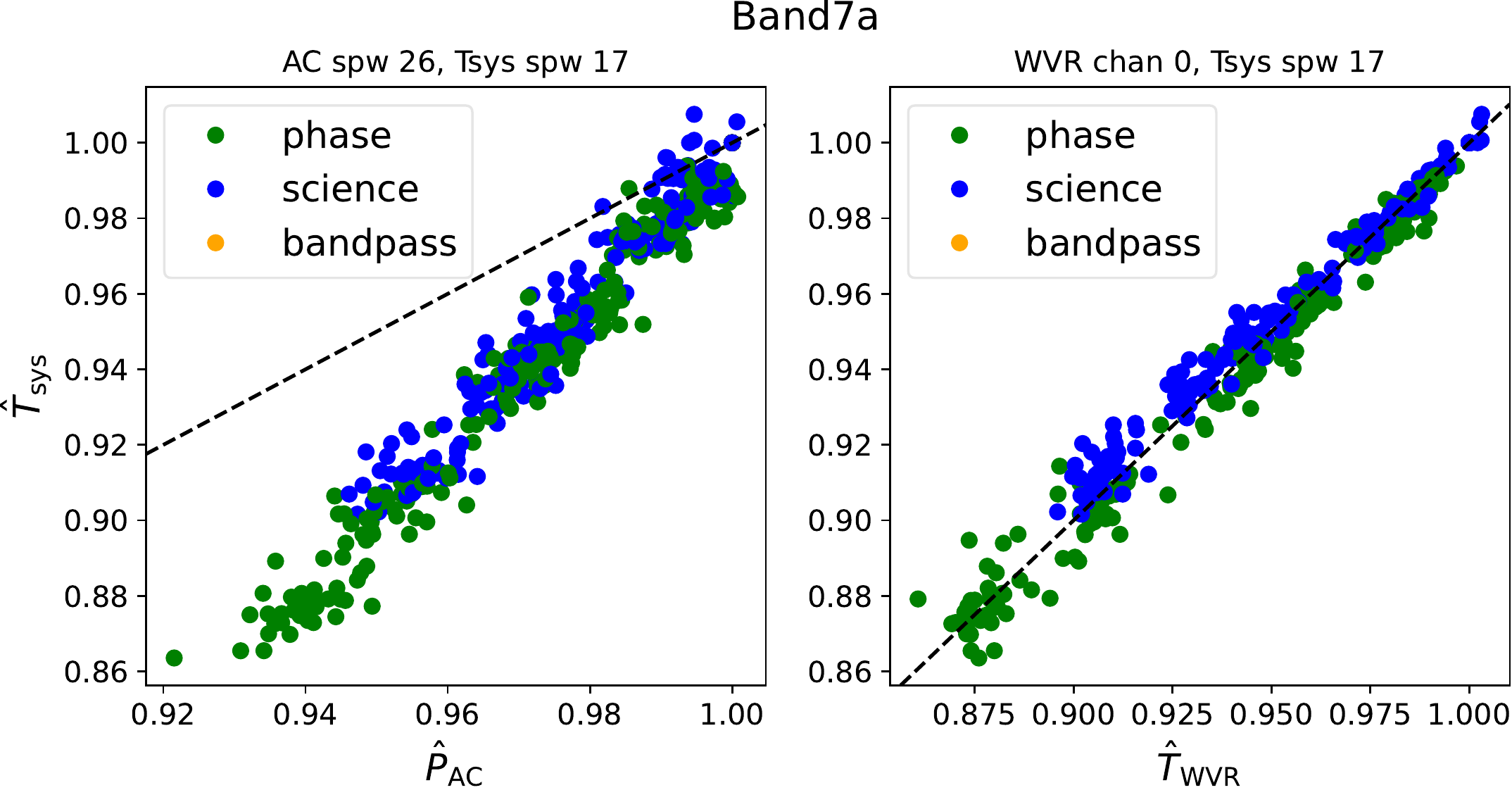}{0.9\linewidth}{}
} 
	\vspace{-2\baselineskip}
	\caption{The correlation between the \TsysNorm and matched \TwvrNorm and normalized $P_{\mathrm{AC}}$ for all the antennas. Both WVR and auto-correlation data is averaged over 10 seconds.}
	\label{fig:AC_correlation}
\end{figure*} 

\begin{figure*}[tbh!]
	\gridline{
		\fig{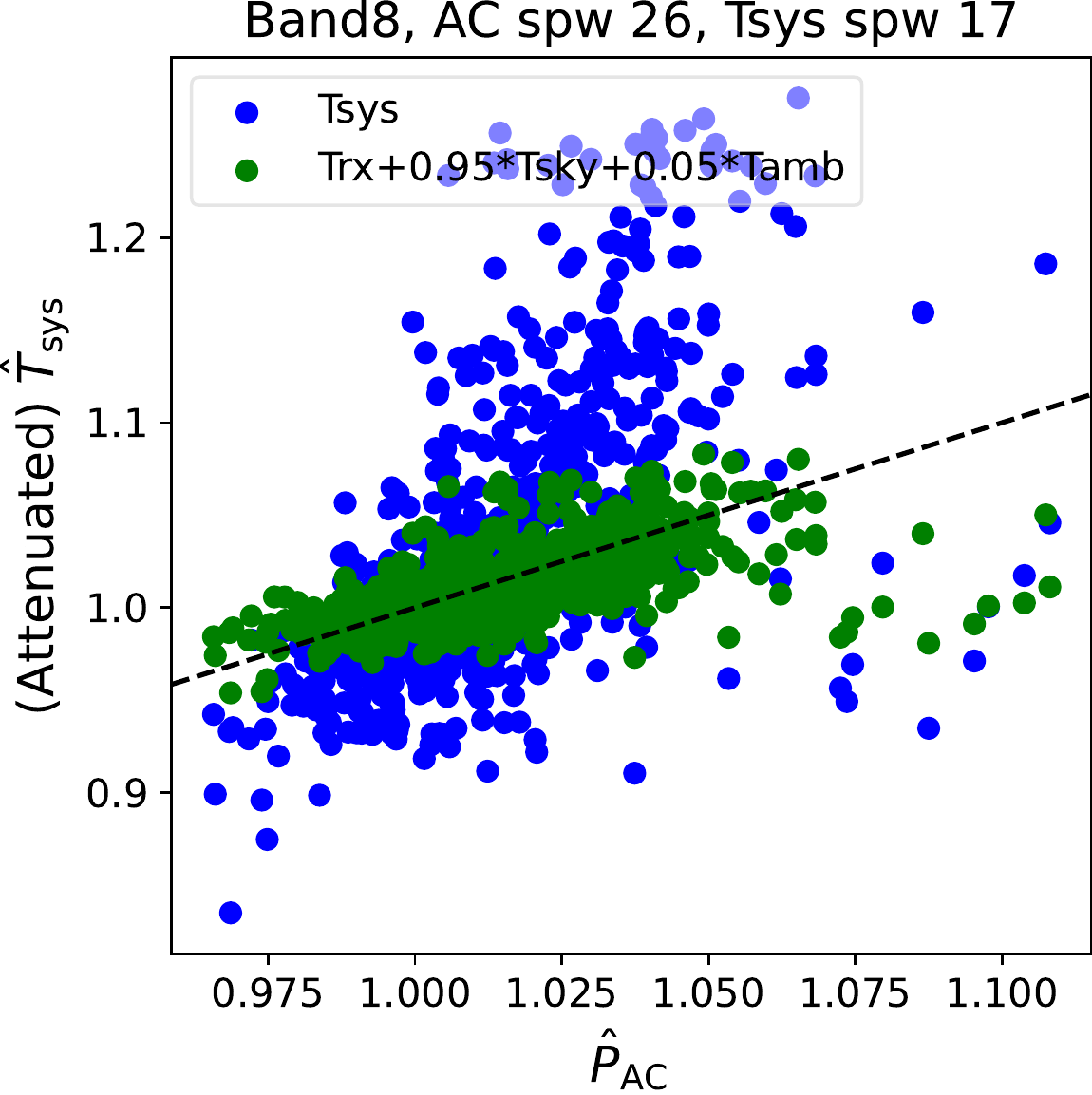}{0.45\linewidth}{}
		\fig{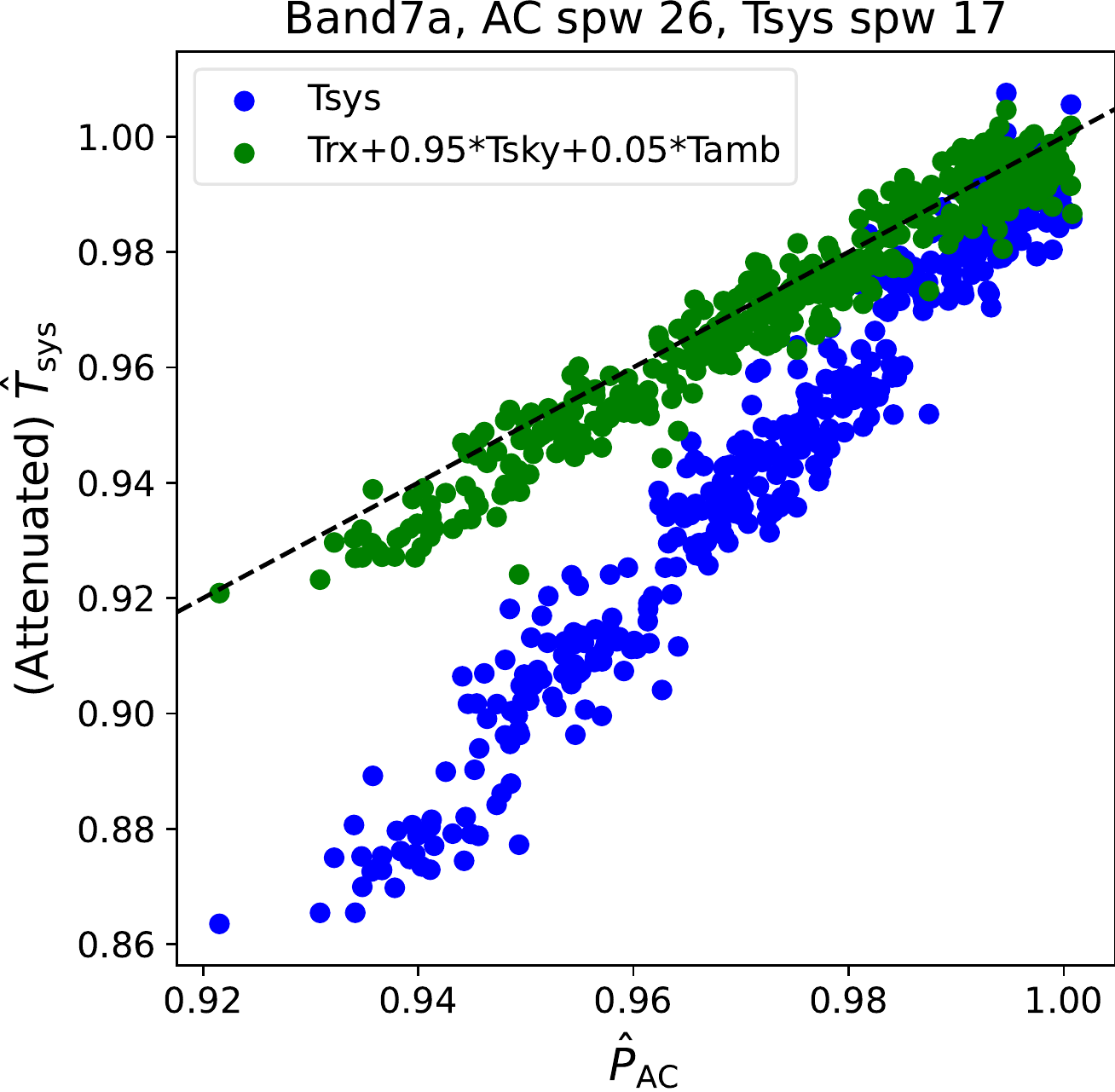}{0.45\linewidth}{}
	} 
	\vspace{-2\baselineskip}
	\caption{The correlation between the normalized \Tsys (blue) and attenuated \Tsys (green) and normalized AC data for the dataset Band8 and Band7a we have for all antennas of one spectral window. As we can see, the attenuated \Tsys follows the 1-to-1 correlation with AC data, which proves equation \ref{eq:Tsys_attenuated} to be right.  }
	\label{fig:AC_correlation_corr}
\end{figure*} 

\begin{figure*}[tbh!]
	\gridline{
		\fig{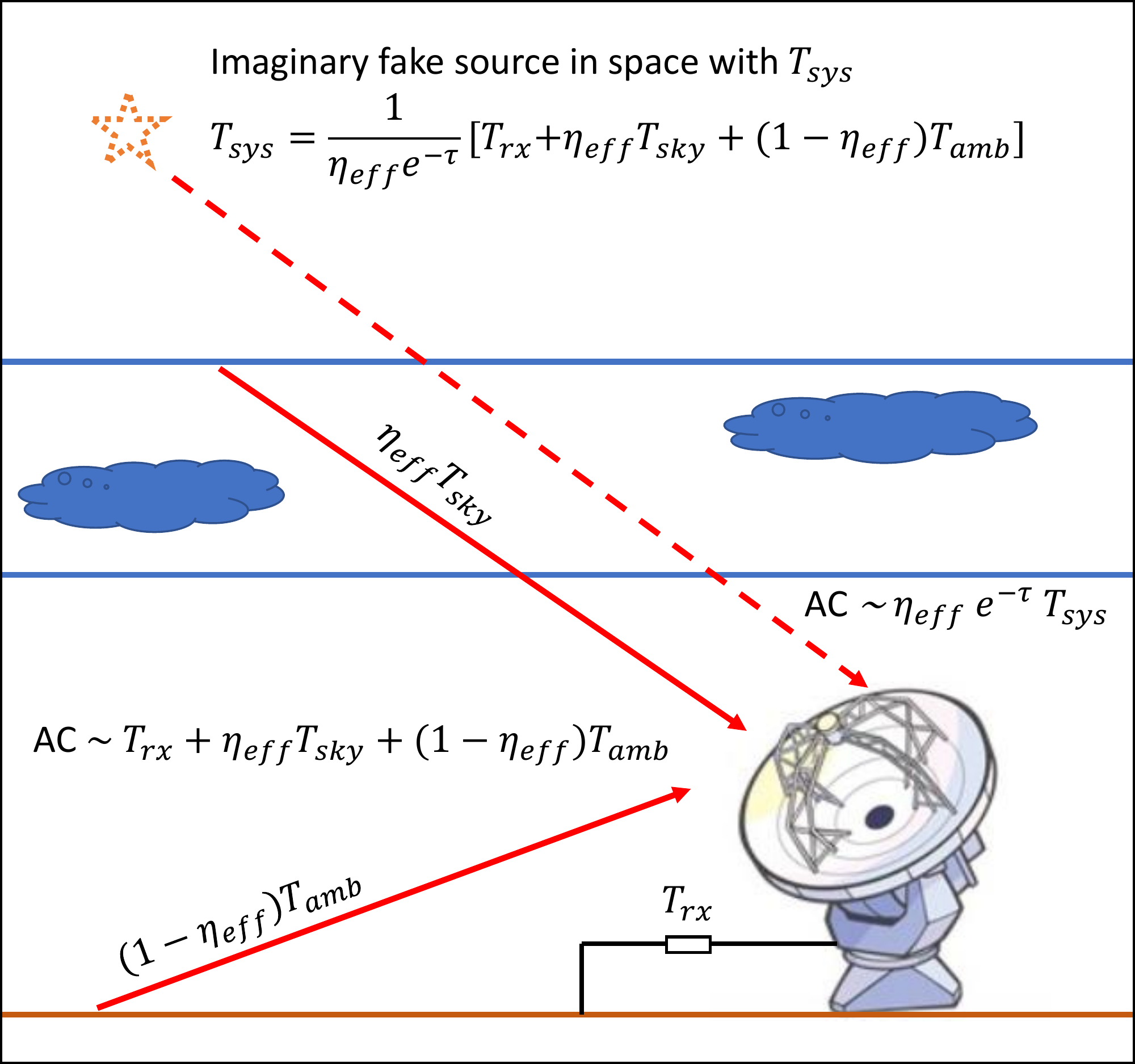}{0.6\linewidth}{}
	} 
	\caption{Illustration of the theoretical relationship between the autocorrelation signal AC and \Tsys. AC is proportional to the total power signal received by an antenna, which includes receiver noise (\Trx), sky noise (\Tsky), and thermal noise due to losses and spillover terminating around the ambient temperature ($(1-\eta_{\mathrm{f}}) \times T_{\mathrm{amb}}$). However, \Tsys is corrected for a source outside the atmosphere, and includes an extra term due to the atmospheric attenuation.  }
	\label{fig:AC_diagram}
	\vspace{\baselineskip}
\end{figure*} 

As mentioned in Section \ref{sec:candidate_data}, we can also explore whether to use AC or SQLD data to track \Tsys variation. Since AC and SQLD data are equivalent to one another (see Fig. \ref{fig: WVR_TP_SQLD} right panel), we only need to compare \Tsys with one of the two quantities. We use AC data for comparison since the data size is much smaller. We use a similar method to normalize the AC data and compare it with the normalized \Tsys. Fig. \ref{fig:AC_correlation} shows the comparison between \Tsys versus AC and \Tsys versus WVR correlation for dataset Band8 and Band7a. We can see that the AC data also has a tight correlation with \Tsys for dataset Band7. However, the correlation does not work well for dataset Band8 with a large scatter. Therefore, we cannot just fit the relation to several Atm-cal scans to calculate the continuous \Tsys with good precision. 

The other thing we find is that AC and \Tsys do not follow the proportional correlation as might be expected. The major reason is that AC data track the total signal received after the atmosphere attenuation while \Tsys tracks the total signal before it comes through the atmosphere, as shown in the diagram in Fig. \ref{fig:AC_diagram}. The AC data is directly proportional to the total signal received by the antenna, which is mainly comprised of emission from the sky ($\eta_{\mathrm{f}} T_{\mathrm{sky}}$), the receiver itself (\Trx), and other fixed losses terminating at ground ($(1-\eta_{\mathrm{f}}) \times T_{\mathrm{amb}}$). However, based on Eq. \ref{eq:Tsys_def}, \Tsys is not directly proportional to these 3 components added together. Instead, \Tsys can be thought of as brightness temperature of a fake source in space that generates a signal equal to the 3 components added together after atmosphere attenuation. In other words, for a single band setting, 
\begin{equation}
\label{eq:Tsys_attenuated}
\begin{split}
P_{\mathrm{AC}} & \propto \eta_{\mathrm{f}} e^{-\tau_0 \sec z} T_{\mathrm{sys}} \\
&\approx T_{\mathrm{rx}} + \eta_{\mathrm{f}} T_{\mathrm{sky}} + (1-\eta_{\mathrm{f}}) \times T_{\mathrm{amb}} 
\end{split}
\end{equation}
We call the right side of the equation attenuated \Tsys. 
We also normalize the attenuated \Tsys the same way as we do for the original \Tsys and compare it with normalized AC data. The comparison between \Tsys and attenuated \Tsys versus AC correlation is shown in Fig. \ref{fig:AC_correlation_corr}. As we can see, the normalized attenuated \Tsys follows the 1-to-1 relation with normalized AC as suggested by Eq. \ref{eq:Tsys_attenuated}. If we rearrange Eq. \ref{eq:Tsys_attenuated}, it becomes
\begin{equation}
\label{eq:PAC_Tsys}
\begin{split}
T_{\mathrm{sys}} \propto \frac{P_{\mathrm{AC}}}{e^{-\tau_{\mathrm{sky}}}} \approx \frac{P_{\mathrm{AC}}}{1-T_{\mathrm{sky}}/T_{\mathrm{amb}}} 
\end{split}
\end{equation}
Therefore, even though we have the AC data, we still need a method to continuously determine \Tsky or $\tau_{\mathrm{sky}}$ to obtain \Tsys. A possible work-around is to use AC data to track \Tsky first by combining Eq. \ref{eq:Tsys_simplified}, \ref{eq:Tsky_Tamb} and \ref{eq:PAC_Tsys} as we can generally assume $T_{\mathrm{amb}}$ and \Trx to be constant. This technique has been applied to correct \Tsys values in \citet{Agliozzo_2017}. However, in our case to extrapolate continuous \Tsys,  we need to note that the AC and SQLD data also suffer from gain drift and gain step changes between scans, as noted above and seen in Fig. \ref{fig: WVR_TP_SQLD}.  

\section{Applying Continuous \Tsys to the Calibration}

In Section \ref{sec:WVR_Tsys}, we demonstrated the viability to use WVR data to track \Tsys continuously. In this section, we apply the extrapolated continuous \Tsys in calibration to test whether our new method for measuring \Tsys works. We calibrate each dataset with the original \Tsys table, the new continuous \Tsys table extrapolated using all Atm-cal scans and that using just 4 Atm-cal scans with CASA package. We then make images from data calibrated using these 3 different methods and see if the measured fluxes for the same target are more consistent with each other using our new methods. The detailed description of the scripts we use for the data processing
can be found at \url{https://github.com/heh15/ALMA_intern_Tsys.git}. 

\subsection{Creation of \Tsys Table}
\label{sec:Tsys_table_generate}

In this subsection we discuss how we construct the new \Tsys table used for the calibration. We note that the original \Tsys table used for calibration is a spectrum with two polarizations. Recording all the extrapolated \Tsys spectra in one big table would take a lot of disk space. Based on our check of the \Tsys spectrum plots generated using the original calibration script, the shape of the \Tsys spectrum of each spectral window does not vary much as a function of time. Therefore, we can just record the initial \Tsys for each observing target in the \Tsys table and record the ratio of the extrapolated \Tsys relative to the initial \Tsys into an amplitude gain table. In this case, the two tables we provide for \Tsys calibration are
\begin{equation}
\label{eq:Tsys_cal_table}
\begin{split}
T_{\mathrm{sys}} (t,\nu) &= T_{\mathrm{sys, obs}}(\mathrm{1^{st}}, \nu) \\
G (t) &= \sqrt{1 \bigg/
	 \left[\frac{T_{\mathrm{sys}}(t)}{T_{\mathrm{sys}}(\mathrm{1^{st}})}\right]_{\mathrm{fit}}}
\end{split} 
\end{equation} 
where $T_{\mathrm{sys}} (t,\nu)$ is the recorded \Tsys spectrum as a function of time $t$ and frequency $\nu$, and $G$ is the derived gain as a function of $t$,  $T_{\mathrm{sys, obs}}(\mathrm{1^{st}}, \nu)$ is the first \Tsys spectrum measured for each type of observation of given antenna and spectral window and $\left[\frac{T_{\mathrm{sys}}}{T_{\mathrm{sys}}(\mathrm{1^{st}})}\right]_{\mathrm{fit}}$ is the extrapolated normalized \Tsys from the fitting. We note that $G$ is not directly equal to the \TsysNorm values. This is due to the different methods that CASA uses to handle \Tsys and gain table. For each baseline, the correlated amplitude is 
\begin{equation}
\begin{split}
S(i,j) &\propto \sqrt{T_{\mathrm{sys}}(i)T_{\mathrm{sys}}(j)} \propto \frac{1}{G(i) G(j)}
\end{split}
\end{equation}
Therefore, the $G$ is written so that it can be properly translated to the variation in \Tsys. 

In Section \ref{sec:fewer_Tsys}, we tested using only 4 \Tsys measurements to fit the linear relation between \TsysNorm and \TwvrNorm. We saw that the difference between this method and using all \Tsys measurements is small. However, we still need to quantify if the small difference in the linear fits makes much difference in the measured flux of the image product. In this case, we also apply Eq. \ref{eq:Tsys_cal_table} to create the alternative \Tsys table with the fitting relation derived from 4 Atm-cal scans. 

\begin{figure*}[]
	\gridline{
		\fig{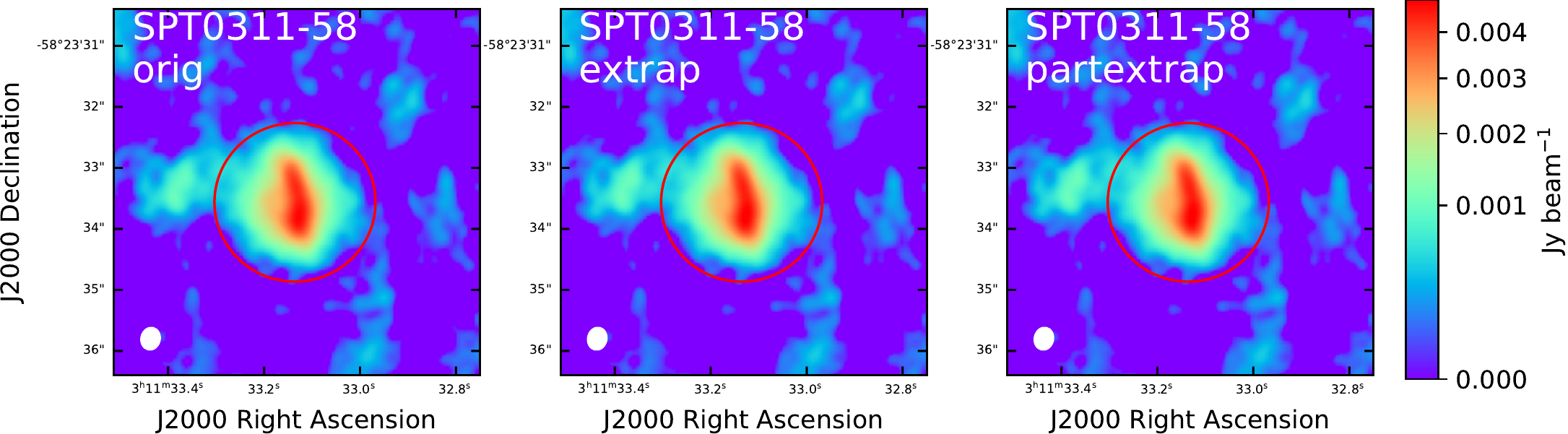}{\linewidth}{}
	}
	\vspace{-1\baselineskip}
	\gridline{
		\fig{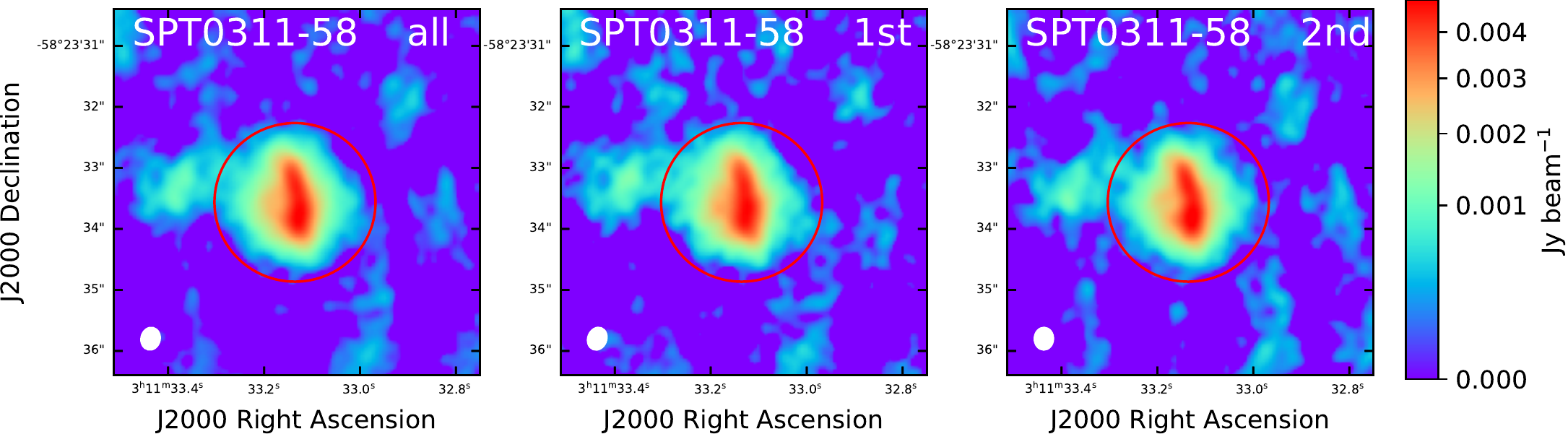}{\linewidth}{}
	}
	\vspace{-1\baselineskip}
	\caption{(Top) dirty images made using originally measured \Tsys (orig), extrapolated continuous \Tsys with all Atm-cal scans (extrap) and that with just 4 Atm-cal scans (partextrap) for dataset Band8 of SPT0311-58. The red circle is the aperture used for flux measurements. The fluxes for these 3 images are 0.0425, 0.0429 and 0.0407 Jy. (Bottom) dirty images made using alternative continuous \Tsys table derived from fitting all Atm-cal scans. The 3 columns are images made using all, 1st half and 2nd half of the science scans. The fluxes for these 3 images are 0.0429, 0.0432 and 0.0428 Jy. }
	\label{fig:Arp220_images}
\end{figure*}

\begin{figure*}
	\gridline{
		\fig{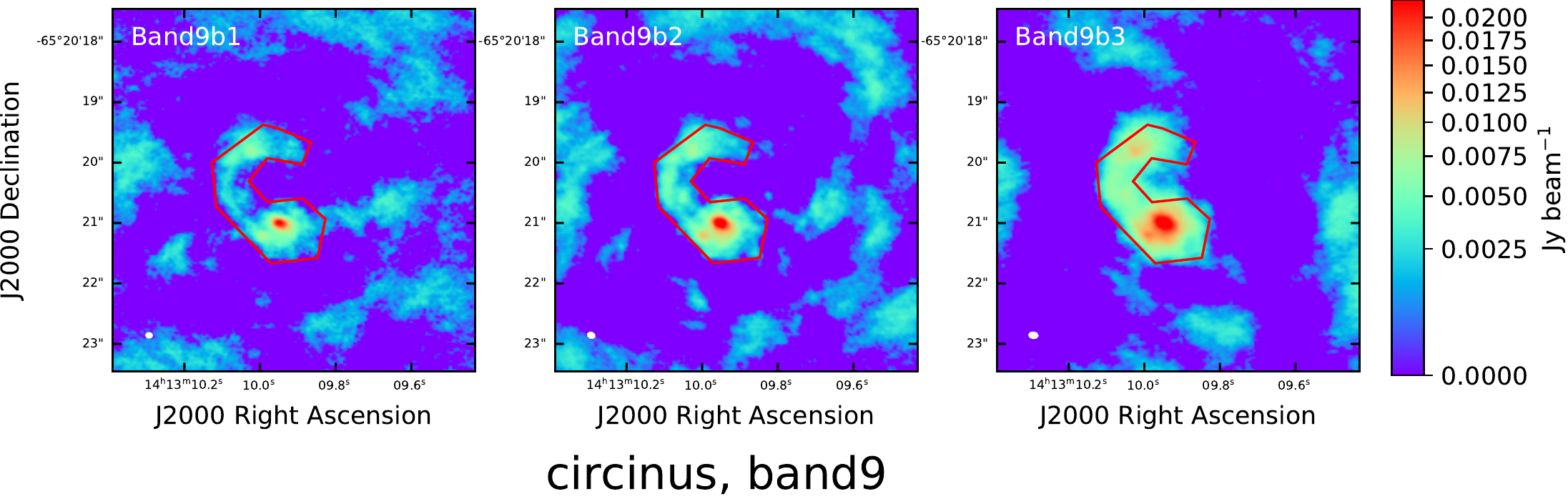}{\linewidth}{}
	}
	\vspace{-1\baselineskip}
	\caption{Dirty images made using alternative continuous \Tsys table using all Atm-cal scans for the 3 data sets in the Band 9 project 2019.1.00013.S. The red polygons are apertures used to measure the flux. The fluxes for these 3 images are 0.875, 1.0 and 1.35 Jy. }
	\label{fig:circinus_band9}	
\end{figure*}

\subsection{Calibrating and Imaging the Data}

After we create the \Tsys table, we then apply the continuous \Tsys in calibration. The calibration script we use is generated from the command \texttt{es.generateReducScript()} \citep{Petry_2014}. We then modify the script to use the alternative \Tsys and amplitude gain tables created (see details in \url{https://github.com/heh15/ALMA_intern_Tsys.git}). We also run the original calibration script to calibrate the data with the original discrete \Tsys method for comparison. 

After calibration, we then proceed with making continuum images. We generally adopt the default settings using the command \texttt{tclean}. We set the \texttt{robust} parameter to be 2.0 instead of the default 0.5 to maximize the sensitivity and hence flux accuracy. We also set the number of iterations to be 0 to only make the dirty image. This process will reduce any effects from \texttt{tclean} itself when making comparisons between fluxes from different calibration datasets. For projects with multiple datasets, we directly compare the measured fluxes from different datasets to see if they are consistent with each other. For projects with just one measurement set, we further make images using just the first half or the second half of the science scans and then compare fluxes among these 3 images. The top panel in Fig. \ref{fig:Arp220_images} shows an example of images made with the 3 different methods for dataset Band8. We can see that the structure of these images looks almost the same. This is what we expect since \Tsys should only affect the intensity scale of the final image. The bottom panel of Fig. \ref{fig:Arp220_images} shows the images of the same target made with all scans or just half of the scans using the new  method to extrapolate \Tsys from all Atm-cal scans. We can see the structures of these images are also almost the same, which is also what we expect as images from the same dataset should have similar $uv$ coverage. On the other hand, if we compare images made from different datasets, we can see there is a larger difference in image structure. This is most obvious in the Band 9 project 2019.1.00013.S shown in Fig. \ref{fig:circinus_band9}. The beam sizes for the 3 different images are also significantly different from each other. For dataset Band7b, all images have similar beam shapes so differences among the image structures are not that significant. 

Once the images are made, we draw an aperture around the central point source to measure the flux. Since the change in \Tsys does not change the structure of the continuum image, we can compare the fluxes measured from the same aperture for the same target (as long as the aperture is not missing any flux, or there is no decorrelation -- as decorrelation will reduce the total flux potentially). The apertures we used to measure the fluxes are shown in Fig. \ref{fig:Arp220_images}, \ref{fig:circinus_band9} and \ref{fig:images_apertures}. Note that as well as changes in \Tsys, changes in phase decorrelation during a single observation and between observations may also affect the measured fluxes, and account for some of the scatter in fluxes in Table \ref{tab:flux_single}, \ref{tab:flux_band7_multi} and \ref{tab:flux_band9_multi}. However, we assume this effect is the same in all reductions, independent of the \Tsys calibration method.

\begin{table*}[tbh!]
	\centering
	\caption{Flux measured for project with only one dataset}
	\label{tab:flux_single}
	\begin{threeparttable}
		\begin{tabularx}{0.95\textwidth}{cccccccc}
			\hline
			Data Label & Target      & Scans used & \multicolumn{3}{c}{Flux (Jy)}                & Meas. Err. (Jy) & Beam (") \\
			&             &            & Tsys\_orig & Tsys\_extrap & Tsys\_partextrap &                 &          \\
			(1)        & (2)         & (3)        & (4)        & (5)          & (6)              &         (7)        &    (8)      \\ \hline
			Band10     & Arp 220     & all        & 7.169      & 7.146        & 7.187            &         0.39        &      0.5 x 0.47    \\
			&             & 1st half   & 7.1638     & 7.135        & 7.171            &       0.42          &   0.52 x 0.45       \\
			&             & 2nd half   & 7.1077     & 7.09         & 7.133            &        0.4         &    0.48 x 0.48      \\
			&             & MAX. DIFF$^a$  & 0.06       & 0.056        & 0.054            &                 &          \\ \hline
			Band9a     & IRAS16293-B & all        & 10.43      & 10.36        & 10.324           &      0.34           &     0.34 x 0.27     \\
			&             & 1st half   & 10.4       & 10.32        & 10.32            &     0.36            &   0.35 x 0.27       \\
			&             & 2nd half   & 10.47      & 10.4         & 10.34            &      0.38           &    0.34 x 0.26      \\
			&             & MAX. DIFF. & 0.07       & 0.04         & 0.02             &                 &          \\ \hline
			Band8      & SPT0311-58  & all        & 0.0425     & 0.0429       & 0.0429           &      0.0015                &  0.35 x 0.3        \\
			&             & 1st half   & 0.0453     & 0.0432       & 0.0433           &    0.0019        &   0.36 x 0.29       \\
			&             & 2nd half   & 0.0407     & 0.0428       & 0.0429           &      0.0017           &     0.35 x 0.3     \\
			&             & MAX. DIFF  & 0.0046     & 0.0003       & 0.0003           &                 &          \\ \hline
			Band7a     & HT-Lup      & all        & 0.173      & 0.178        & 0.178            &           0.0038      &  0.22 x 0.12        \\
			&             & 1st half   & 0.169      & 0.175        & 0.175            &         0.0044        &      0.22 x 0.12    \\
			&             & 2nd half   & 0.175      & 0.18         & 0.18             &        0.0045         &      0.22 x 0.12    \\
			&             & MAX. DIFF. & 0.006      & 0.005        & 0.005            &                 &          \\ \hline
		\end{tabularx}
		\begin{tablenotes}[flushleft]
			\item 
			\textbf{Columns:} (1) The label of each dataset (see Table \ref{tab:data_summary}) (2) The target name. (3) The science scans used to make images. (4) Fluxes of the source with images made using original calibration script. (5) Fluxes of the source with images made using modified script with alternative \Tsys table. The \Tsys is extrapolated based on all Atm-cal scans. (6) The images made with modified script but \Tsys is extrapolated from 4 Atm-cal scans. (7) The measured flux errors for the images made with original \Tsys table (8) The beam size of images using original \Tsys table. 
			\item \textbf{Rows:} a. The maximal differences for fluxes at each column. 
		\end{tablenotes}
	\end{threeparttable}
\end{table*}

\begin{table*}[tbh!] 
	\centering
	\caption{Flux measured for data for Band7b project (AS205A)}
	\label{tab:flux_band7_multi}
	\begin{threeparttable}
		\begin{tabular}{cccccc}
			\hline
			Dataset Label & \multicolumn{3}{c}{Flux (Jy)}                & Meas. Err. (Jy) & Beam (") \\
			& Tsys\_orig & Tsys\_extrap & Tsys\_partextrap &                 &          \\
			(1)           & (2)        & (3)          & (4)              & (5)             & (6)      \\ \hline
			Band7b1       & 0.7699     & 0.759        & 0.7594           &      0.031           &     1.13 x 0.79     \\
			Band7b2       & 0.713      & 0.7146       & 0.7129           &      0.029           &     0.96 x 0.77     \\
			Band7b3*      & 1.001      & 1.005        & 1.003            &       0.043          &     0.68 x 0.53     \\
			Band7b4       & 0.747      & 0.7378       & 0.7359           &     0.028            &  1.03 x 0.78        \\
			Band7b5*      & 0.556      & 0.58         & 0.58             &    0.029             &   1.31 x 0.72       \\
			Band7b6       & 0.749      & 0.7534       & 0.7533           &   0.03              &    0.91 x 0.08      \\
			Band7b7       & 0.7355     & 0.7347       & 0.7348           &      0.029           &     1.01 x 0.81     \\
			Band7b8       & 0.768      & 0.737        & 0.7372           &      0.029           &      1.24 x 0.79    \\
			AVG.$^a$          & 0.7471     & 0.7394       & 0.7389           &                 &          \\
			STD.$^b$          & 0.0194     & 0.0143       & 0.0149           &                 &          \\ \hline
		\end{tabular}
		\begin{tablenotes}
			\item \textbf{Columns:} (1) The label for each dataset (see Table \ref{tab:data_summary}). (2) Flux of the data set using the original \Tsys table. (3) Flux of the data set using alternative continuous \Tsys table with the linear relation fitted using all Atm-cal scans. (4) Flux of the data set using continuous \Tsys table with the linear relation fitted using part of Atm-cal scans. (5) Measured flux errors using original \Tsys table. (6) Beams of the image using original \Tsys table. 
			\item \textbf{Rows: } a. The average value for each column. b. The standard deviation for each column. 
			\item \textbf{Notes:} * denotes data with unusual fluxes. Fluxes from these data are not included in the calculation of the average and standard deviation value.  
		\end{tablenotes}
	\end{threeparttable}
\end{table*}

\begin{table*}[tbh!] 
	\centering
	\caption{Flux measured for data for dataset Band9b (Circinus)}
	\label{tab:flux_band9_multi}
	\begin{threeparttable}
		\begin{tabular}{cccccc}
			\hline
			Dataset Label & \multicolumn{3}{c}{Flux (Jy)}                & Meas. Err. (Jy) & Beam (") \\
			& Tsys\_orig & Tsys\_extrap & Tsys\_partextrap &                 &          \\
			(1)           & (2)        & (3)          & (4)              & (5)             & (6)      \\ \hline
			Band9b1       & 0.85       & 0.875        & 0.849            &       0.018          &   0.086 x 0.065       \\
			Band9b2       & 0.96       & 1            & 0.983            &      0.021           &   0.104 x 0.077       \\
			Band9b3       & 1.27       & 1.35         & 1.34             &     0.016            &     0.131 x 0.085     \\
			AVG.          & 1.027      & 1.075        & 1.057            &                 &          \\
			STD.          & 0.178      & 0.201        & 0.207            &                 &          \\ \hline
		\end{tabular}
		\begin{tablenotes}
			\item \textbf{Columns:} (1) The label for each dataset (see Table \ref{tab:data_summary}).  (2) Flux of the data set using the original \Tsys table. (3) Flux of the data set using alternative continuous \Tsys table with the linear relation fitted using all Atm-cal scans. (4) Flux of the data set using continuous \Tsys table with the linear relation fitted using part of Atm-cal scans. 
			\item \textbf{Rows: } a. The average value for each column. b. The standard deviation for each column. 
		\end{tablenotes}
	\end{threeparttable}   
\end{table*}  

\begin{figure*}[tbh!]
	\gridline{
		\fig{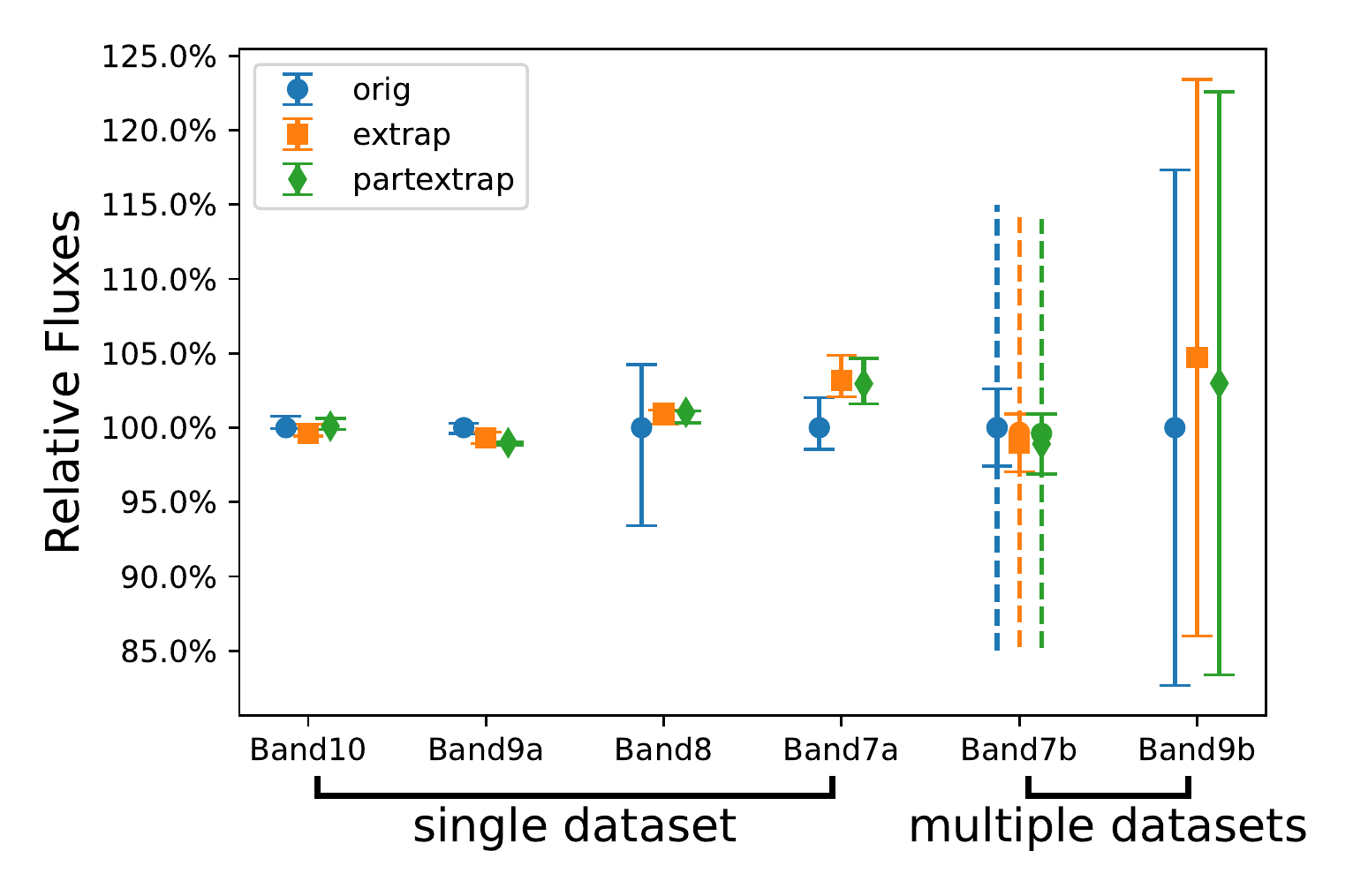}{0.7\linewidth}{}
	}
	\vspace{-2\baselineskip}
	\caption{Comparison of fluxes of images made using the original \Tsys (orig), continuous \Tsys extrapolated from all Atm-cal scans (extrap) and that extrapolated from 4 Atm-cal scans (partextrap). The vertical axis are the ratio of fluxes to the fluxes using original method. The first 4 projects contain single dataset while the last two projects contain multiple datasets.	The dashed line for Band7b project shows the error including the two datasets that has abnormal flux values (see Table .\ref{tab:flux_band7_multi}). 
		We can see the flux consistency is generally better for most of the projects with smaller uncertainties. See description in Section \ref{sec:flux_compare}.  }	
	\label{fig:flux_comparison}
\end{figure*}

\subsection{Flux Comparison}
\label{sec:flux_compare}

The measured fluxes are recorded in Table \ref{tab:flux_single}, \ref{tab:flux_band7_multi} and \ref{tab:flux_band9_multi}. The flux uncertainty can be separated into two parts, the measurement error and the calibration error. The measurement error is calculated as 
\begin{equation}
\mathrm{Meas. Err} = \mathrm{rms} \sqrt{N_{\mathrm{beam}}}
\end{equation}
where the rms is the measured noise of the image and $N_{\mathrm{beam}}$ is the number of beams across the aperture used to measure the flux. On the other hand, our alternative method to measure \Tsys should mainly work on reducing the calibration error. To test if our new method improves the flux calibration accuracy, we need to quantify the calibration error for each method we use. For projects with a single dataset, we compare fluxes of images made with 1st half, 2nd half and all scans and calculate the maximal difference between the 3 flux values as the calibration error. For projects with multiple datasets, the calibration error is calculated as the standard deviation of fluxes of the different datasets. Both measurement and calibration errors are recorded in Table \ref{tab:flux_single}, \ref{tab:flux_band7_multi} and \ref{tab:flux_band9_multi}.

To further compare how our methods work for datasets from different frequency bands, we normalize the flux values to the flux value using the original discrete \Tsys calibration method with all Atm-cal scans. The relative uncertainties are calculated as the calibration error divided by the flux value for each method using all Atm-cal scans. The comparison is shown in Fig. \ref{fig:flux_comparison}. We can see that fluxes using the original \Tsys table do not differ significantly from the fluxes using alternative \Tsys and gain tables, with maximal differences smaller than 5\%. We also demonstrate that the extrapolated \Tsys using 4 Atm-cal scans gives us fluxes that are almost the same as when using all Atm-cal scans, which proves it is viable to significantly reduce the number of \Tsys measurements by using WVR-tracked \Tsys. Furthermore, it seems for most of the datasets, our new methods give better flux consistency, especially for dataset Band8 which brings down the flux calibration uncertainty contribution due to \Tsys variability from $\sim$ 10\% to 0.7\%. As we have shown in Section \ref{subsec: Tsys_Twvr_ATM}, the \Tsys variation for this dataset is mainly driven by the PWV variation at short time-scale. In this case, the discrete \Tsys measurements poorly sampled the fluctuations of the real \Tsys (Fig. \ref{fig:Tsys_WVR_extrap}). Our new method instead catches the variation in \Tsys between the discrete ATM calibrations, and thus keeps the flux consistent. 

This method also works for dataset Band7b with multiple datasets for which relative flux uncertainties reduce from 2.5\% to 1.9\%. The only project that gives us larger flux uncertainties using our new methods is the dataset Band9b. For this project, the uncertainties using all 3 methods are quite large ($\sim$ 15\%). The large uncertainties are probably due to different uv-coverages, the complex target structure, the relatively high phase noise and maybe the time variability of the flux calibrator in these datasets. This is also a tricky dataset for which the linear fitting does not work as well as for other data sets. In our future work, we will explore if the larger uncertainty is caused by imperfect fitting of the \TsysNorm versus \TwvrNorm relation.

\subsection{Additional considerations for the continuous \Tsys method}
\label{subsec:additional}

For some targets, the source brightness temperature in single-dish measurements can be significant compared with \Tsys. For example, this may occur for bright galactic targets in 12CO, bright masers, or for some Solar System objects in continuum. As the widths of galactic spectral lines are generally negligible ($<$ 1\%) compared with the normal bandwidth (2GHz) used to measure \Tsys, and the continuous \Tsys method uses a spectrally-averaged broad-band \Tsys, then the effect of bright lines in such cases will be negligible. But for very bright continuum sources such as planets, the spectrally-averaged \Tsys will potentially be affected by the target brightness. However, the beam of the WVR unit on each antenna is offset from the optical axes of the receiver beams by several arc minutes (depending on the receiver band in use - see ALMA Technical handbook); this means that the WVRs are not pointing to the science target, and in general will not be affected by its strong continuum.
Additionally, it has recently become apparent that the method used by ALMA to measure \Tsys, using off-source data along with the normalisation of the visibilities using the autocorrelation, introduces a calibration error for bright sources\footnote{https://help.almascience.org/kb/articles/what-are-the-amplitude-calibration-issues-caused-by-alma-s-normalization-strategy}\footnote{https://help.almascience.org/kb/articles/what-errors-could-originate-from-the-correlator-spectral-normalization-and-tsys-calibration}. The planned change is to measure \Tsys on-source. Again, this should not significantly affect the continuous \Tsys method, for reasons given above. However, we need to note that this technique cannot be applied to solar observing because \Twvr from all WVR channels will be heavily saturated.

For spectral lines, an assumption is made that \Tsys is mainly affected by PWV, and the correlation of \Tsys with PWV uses \Tsys averaged over the spectral window. This is considered reasonable for continuum and most spectral lines, but for calibration of spectral lines coincident with deep Ozone absorption (e.g. see Fig. \ref{fig:ATM_spectrum}), the correlation will have a slightly different slope and intercept. In general this is considered a second-order effect; for example, a line exactly coincident with the strong O$_3$ peak at 428.8GHz in Fig. \ref{fig:ATM_spectrum}, the error in the correction of \Tsys based on the PWV would be $\sim \pm$ 3\%. A future improvement might be correct the data spectrally rather than using a single channel-averaged value per timestamp. However, this would make the correction table significantly larger (see Section \ref{sec:Tsys_table_generate}). 

An additional use of the continuous \Tsys method could be to correct for the increase in \Tsys due to shadowing of the antennas. On ALMA, the default is that data taken with any slight blocking of the beam from an antenna, for example by a nearby antenna or building, is flagged and removed during data reduction. In general this cannot be corrected for using the gain calibrator amplitude solution, as this is not observed at the same sky location as the target. However, if the corresponding increase in \Tsys due to shadowing is measured continuously, it may be possible to calibrate out some degree of shadowing. Further investigation of this technique should be done.

\section{Conclusions and Future Work}

In this paper, we explore a new method to use continuous datastreams available from WVR monitoring to track the atmospheric opacity and hence \Tsys in mm and submm data. The aim is to improve flux calibration in conditions where the sky opacity is rapidly varying, and to reduce overheads needed for frequent discrete calibration using internal loads. Here we summarize our main conclusions regarding initial tests of this method. 

\begin{itemize}
	\item There is a tight linear correlation between normalized \Tsys and \Twvr, with typical scatter of $\sim$ 1\%. For the worst case of Band 9 data with large \Tsys variations (50\%), the simple linear fit would give us scatter of $\sim$ 4\%, which is due to the non-linearity of the relation at high opacities with large \Tsys variations. Although the exact form of the linear relation varies among different spectral windows and different data sets, we can use as few as 4 Atm-cal scans to determine the slope and intercept of the linear relation, which suggests it is possible to significantly reduce the number of discrete \Tsys measurements during observations, particularly at high frequencies. Furthermore it is not necessary to perform separate calibrations on the phase calibrator and science target, as the continuous \Tsys method is able to track differences in \Tsys between the two. 
	
	\item We have successfully reproduced the observed tight \TsysNorm vs \TwvrNorm correlation using ATM modeling for several datasets. The ATM modeling suggests that changing elevation or PWV will give us \TsysNorm vs \TwvrNorm relation of different slopes when the dry opacity is significant at the observing frequencies. This suggests that we might not get the tight \TsysNorm vs \TwvrNorm correlation at these frequencies if both PWV and elevation varies significantly. A better strategy to track \Tsys, especially in the cases mentioned above, would be to calculate the PWV using the continuous \Twvr measurements from different WVR channel and combine the measured PWV, elevation and \Trx for the dataset to derive the continuous \Tsys based on the ATM modeling.  
	
	\item We apply the continuous \Tsys in calibration and find that it generally gives us more consistent fluxes for the same target.
	For the dataset Band8 which has the largest PWV variation, the flux calibration uncertainty contribution due to \Tsys variability is reduced from 10\% to 0.7\%. The only exception is the dataset Band9b as our new methods give higher flux uncertainties. We suspect part of reasons are due to the imperfect linear fitting of the \Tsys vs \Twvr relation. Since the uv-coverages for the data sets in this project are significantly different, it is hard to confirm this scenario for this data set. 
	
	\item If this method is used for sub-mm observatories such as ALMA, it can reduce the number of \Tsys measurements required for  high-frequency observations from 10 $\sim$ 20 down to 5 (4 \Tsys measurements for the fitting and 1 bandpass \Tsys) or fewer. Assuming each observation block takes $\sim$ 60 mins and each \Tsys measurement takes about 30 -- 40 seconds, it has the potential to save $\sim$ 10\% of observing time for high frequency observing, which is made more valuable as the amount of time in such good conditions is limited. 
\end{itemize}
\  \\
\noindent We thank the referee for thoughtful comments and constructive suggestions, particularly regarding ATM modeling. This paper makes use of the following ALMA data: \\
ADS/JAO.ALMA \#2015.1.00271.S \\
ADS/JAO.ALMA \#2016.1.00744.S \\
ADS/JAO.ALMA \#2018.1.01778.S \\
ADS/JAO.ALMA \#E2E8.1.00003.S \\
ADS/JAO.ALMA \#2018.1.01210.S \\
ADS/JAO.ALMA \#2019.1.00013.S. ALMA is a partnership of ESO (representing its member states), NSF (USA) and NINS (Japan), together with NRC (Canada), MOST and ASIAA (Taiwan), and KASI (Republic of Korea), in cooperation with the Republic of Chile. The Joint ALMA Observatory is operated by ESO, AUI/NRAO and NAOJ. The National Radio Astronomy Observatory is a facility of the National Science Foundation operated under cooperative agreement by Associated Universities, Inc. This research made use of Astropy,\footnote{http://www.astropy.org} a community-developed core Python package for Astronomy \citep{astropy:2013, astropy:2018}.  HH acknowledges the support of NSERC-CREATE NTCO training program. The research of C.D.W. is supported by grants from the Natural Sciences and Engineering Research Council of Canada and the Canada Research Chairs program.

\software{astropy \citep{astropy:2013,astropy:2018}, 
	CASA \citep{McMullin_2007, Emonts_2020, Bean_2022}
}

\bibliography{references}{}
\bibliographystyle{aasjournal}

\restartappendixnumbering
\newpage
\appendix

\section{\Tsys vs \Twvr Relation}
\label{sec:extra_figure}

In this section we show the \Twvr vs \Twvr relation and the extrapolated \Tsys for the rest of data we have. 

\begin{figure*}[htb!]
	\gridline{
		\fig{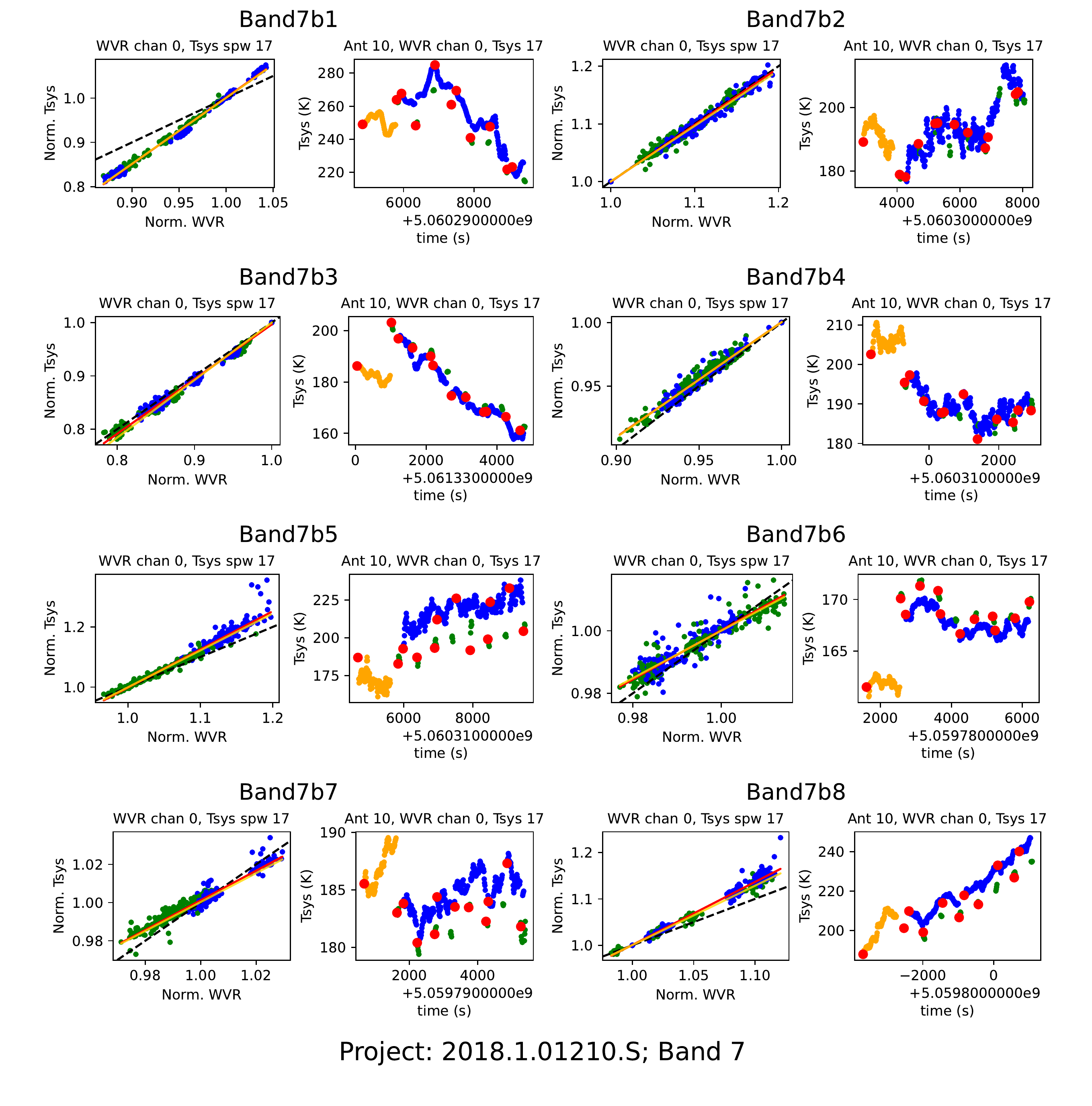}{0.95\linewidth}{}
	} 
	\vspace{-3\baselineskip}
	\caption{The summary of linear relation between \Tsys and \Twvr and comparison between measured and extrapolated \Tsys for measurement sets in project 2018.1.01210.S. For each measurement set, the left subplot shows the linear correlation between normalized \Tsys and \Twvr. The blue and green points are from science and phase scan. The dashed line shows the 1-to-1 relation. The red and golden solid line is the fitted linear relation using data from all Atm-cal scans or just 4 Atm-cal scans. The right plot shows the extrapolated \Tsys based on the fitting relation using all Atm-cal scans. The orange, green and blue points are extrapolated \Tsys for bandpass, phase and science targets. The red points are the original measured \Tsys.  }
	\label{fig:Tsys_WVR_spw17_ant10_summary}
\end{figure*} 

\begin{figure*}[htb!]
	\gridline{
		\fig{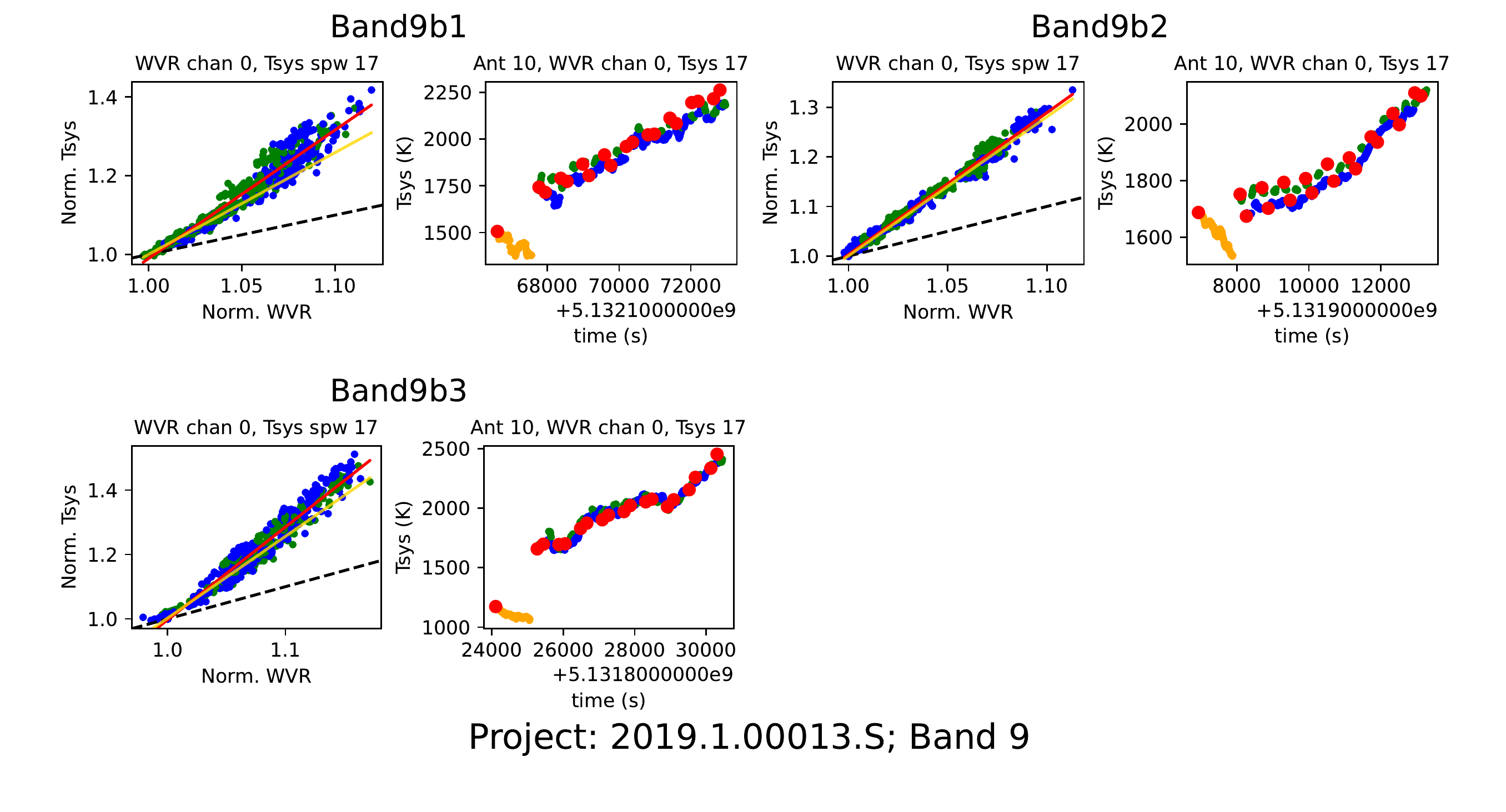}{0.92\linewidth}{}
	} 
	\vspace{-3\baselineskip}
	\caption{The summary of linear relation between \Tsys and \Twvr and comparison between measured and extrapolated \Tsys for measurement sets in project 2019.1.01210.S. See details in Fig. \ref{fig:Tsys_WVR_spw17_ant10_summary}. }
	\label{fig:Tsys_WVR_spw17_ant10_band9}
\end{figure*} 

\begin{figure*}[htb!]
	\gridline{
		\fig{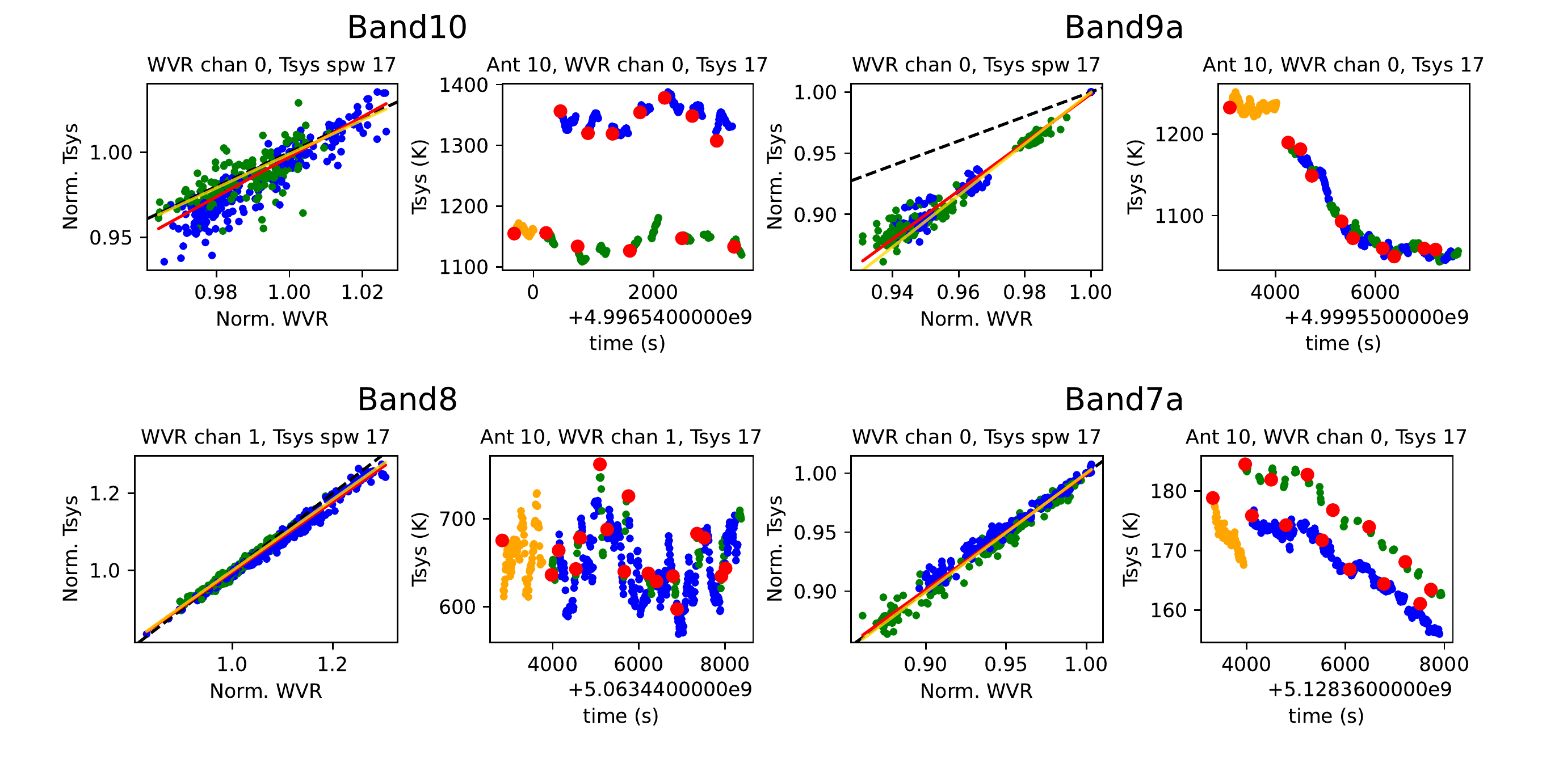}{0.92\linewidth}{}
	} 
	\vspace{-3\baselineskip}
	\caption{The summary of linear relation between \Tsys and \Twvr and comparison between measured and extrapolated \Tsys for the other 4 projects used in this paper with just one measurement set. See details in Fig. \ref{fig:Tsys_WVR_spw17_ant10_summary}. }
	\label{fig:Tsys_WVR_spw17_ant10_single}
\end{figure*} 

\newpage
\section{Dirty images for the rest of targets}

In this section we show the dirty image of the rest of datasets made with extrapolated continuous \Tsys using all Atm-cal scans and the aperture we used to measure the flux. 

\begin{figure*}[tbh!]
	\gridline{
		\leftfig{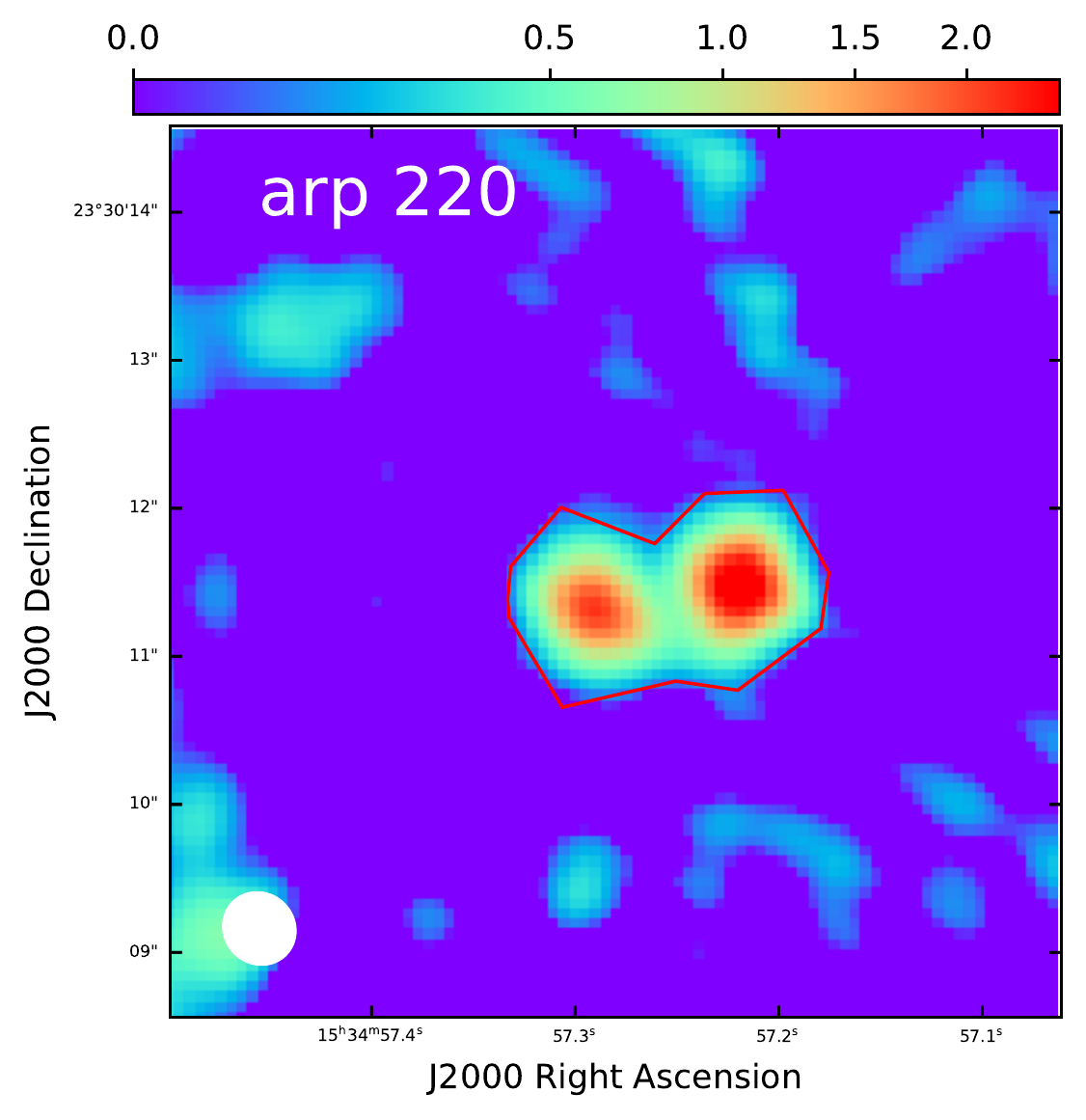}{0.3\linewidth}{}
		\leftfig{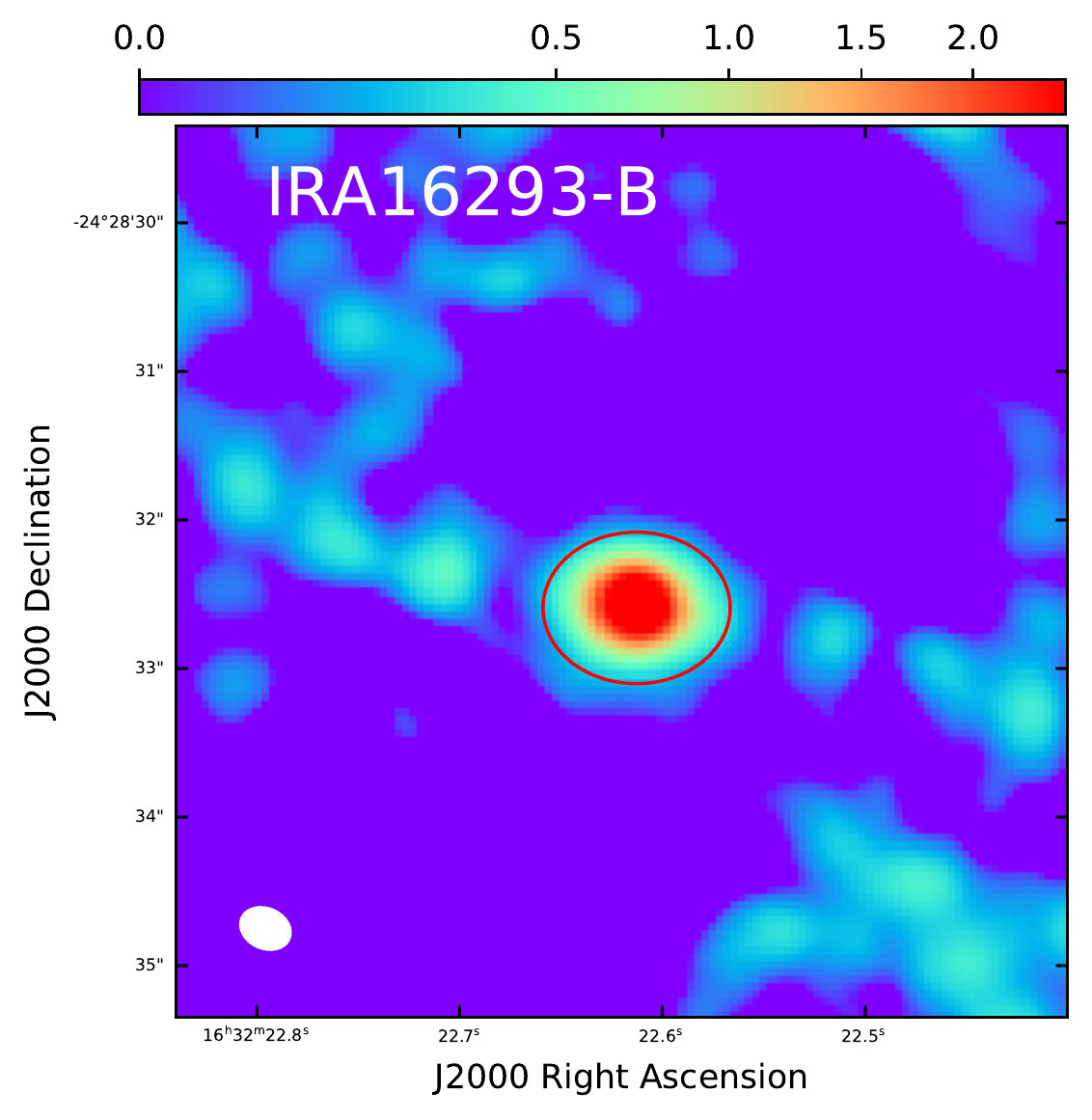}{0.3\linewidth}{}
		\fig{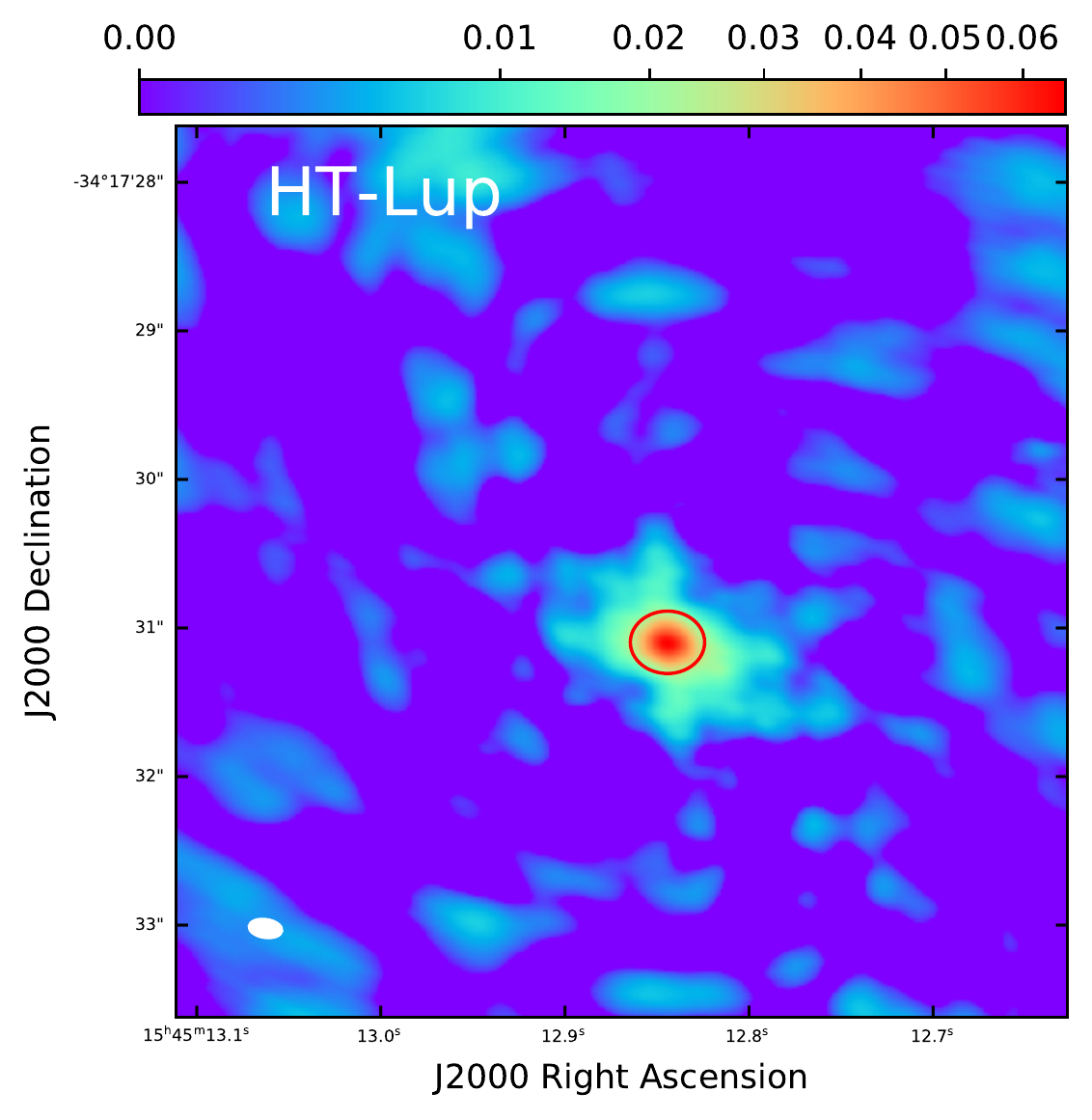}{0.3\linewidth}{}
	}
	\vspace{-1\baselineskip}
	\gridline{
		\leftfig{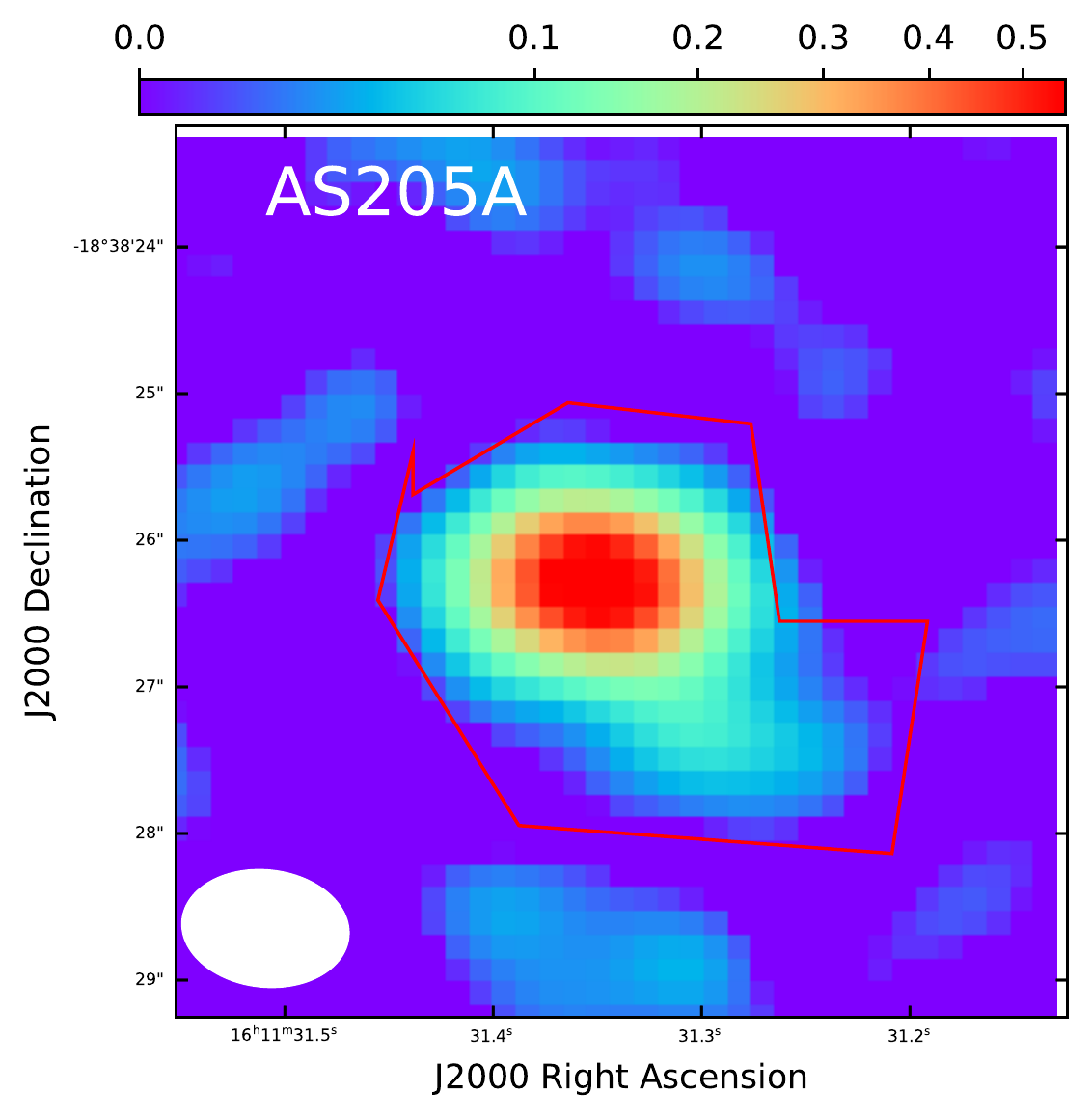}{0.3\linewidth}{}
	}
	\vspace{-2\baselineskip}
	\caption{Dirty images for other sources made using alternative continuous \Tsys table using all Atm-cal scans. For Band 7 project 2018.1.01210.S (AS205A), we show the image made from data uid://A002/Xda1250/X2387 (Band7b1). The red polygons are apertures used to measure the flux.  }
	\label{fig:images_apertures}
\end{figure*}



\end{document}